\documentclass[longauth]{aa}  

\usepackage{graphicx}
\usepackage{txfonts}
\usepackage{natbib}
\bibpunct{(}{)}{;}{a}{}{,}
\usepackage[colorlinks=true, citecolor=blue]{hyperref}
\usepackage{multirow}
\makeatletter
\renewcommand*\aa@pageof{, page \thepage{} of \pageref*{LastPage}}
\makeatother
\usepackage{etoolbox}
\usepackage{array,booktabs}

\begin{document}

   \title{V-LoTSS: The Circularly-Polarised LOFAR Two-metre Sky Survey}
   \titlerunning{V-LoTSS}

   \author{J.~R.~Callingham\inst{1,2}\thanks{Corresponding author \email{jcal@strw.leidenuniv.nl}}
          \and
          T.~W.~Shimwell\inst{2,1}
          \and
          H.~K.~Vedantham\inst{2,3}
          \and
          C.~G.~Bassa\inst{2}
          \and
          S.~P.~O'Sullivan\inst{4}
          \and
          T.~W.~H.~Yiu\inst{2,3}
          \and
          S.~Bloot\inst{2,3}
          \and
          P.~N.~Best\inst{5}
          \and
          M.~J.~Hardcastle\inst{6}
          \and
          M.~Haverkorn\inst{7}
          \and
          R.~D.~Kavanagh\inst{2}
          \and
          L.~Lamy\inst{8,9}
          \and
          B.~J.~S.~Pope\inst{10,11}
          \and
          H.~J.~A.~R\"{o}ttgering\inst{1}
          \and
          D.~J.~Schwarz\inst{12}
          \and
          C.~Tasse\inst{13,14}
          \and
          R.~J.~van Weeren\inst{1}
          \and
          G.~J.~White\inst{15,16}
          \and
          P.~Zarka\inst{8,17}
          \and
          D.~J.~Bomans\inst{18}
          \and
          A.~Bonafede\inst{19,20}
          \and
          M.~Bonato\inst{20}
          \and
          A.~Botteon\inst{19,20}
          \and
          M.~Bruggen\inst{21}
          \and
          K.~T.~Chy\.zy\inst{22}
          \and
          A.~Drabent\inst{23}
          \and
          K.~L.~Emig\inst{24}
          \and
          A.~J.~Gloudemans\inst{1}
          \and
          G.~G\"{u}rkan\inst{23,25}
          \and
          M.~Hajduk\inst{26}
          \and
          D.~N.~Hoang\inst{21}
          \and
          M.~Hoeft\inst{23}
          \and
          M.~Iacobelli\inst{2}
          \and
          M.~Kadler\inst{27}
          \and
          M.~Kunert-Bajraszewska\inst{28}
          \and
          B.~Mingo\inst{29}
          \and
          L.~K.~Morabito\inst{30,31}
          \and
          D.~G.~Nair\inst{32}
          \and
          M.~P\'{e}rez-Torres\inst{33,34,35}
          \and
          T.~P.~Ray\inst{36}
          \and
          C.~J.~Riseley\inst{19}
          \and
          A.~Rowlinson\inst{2,37}
          \and
          A.~Shulevski\inst{1}
          \and
          F.~Sweijen\inst{1,2}
          \and 
          R.~Timmerman\inst{1}
          \and 
          M.~Vaccari\inst{38,39,20}
          \and
          J.~Zheng\inst{12}
          }

   \institute{ Leiden Observatory, Leiden University, PO Box 9513, 2300 RA Leiden, The Netherlands\goodbreak
        \email{jcal@strw.leidenuniv.nl}
        \and
        ASTRON, Netherlands Institute for Radio Astronomy, Oude Hoogeveensedijk 4, 7991 PD Dwingeloo, The Netherlands\goodbreak
        \and
        Kapteyn Astronomical Institute, University of Groningen, PO Box 800, 9700 AV Groningen, The Netherlands\goodbreak
        \and
        School of Physical Sciences and Centre for Astrophysics \& Relativity, Dublin City University, Glasnevin D09 W6Y4, Ireland\goodbreak
        \and
        Institute for Astronomy, University of Edinburgh, Royal Observatory, Blackford Hill, Edinburgh EH9 3HJ, UK\goodbreak
        \and      
        Centre for Astrophysics Research, Department of Physics, Astronomy and Mathematics, University of Hertfordshire, College Lane, Hatfield AL10 9AB, UK\goodbreak
        \and
        Department of Astrophysics/IMAPP, Radboud University, PO Box 9010, 6500 GL Nijmegen, The Netherlands\goodbreak
        \and
        LESIA, Observatoire de Paris, Universit\'{e} PSL, CNRS, Sorbonne Universit\'{e}, Universit\'{e} de Paris,  5 Place Jules Janssen, 92195 Meudon, France\goodbreak
        \and
        Aix Marseille  Universit\'{e}, CNRS, CNES, LAM, Marseille, France, 38 Rue Fr\'{e}d\'{e}ric Joliot Curie, 13013 Marseille, France\goodbreak
        \and
        School of Mathematics and Physics, The University of Queensland, St Lucia, QLD 4072, Australia\goodbreak
        \and
        Centre for Astrophysics, University of Southern Queensland, West Street, Toowoomba, QLD 4350, Australia\goodbreak
        \and 
        Fakult\"{a}t f\"{u}r Physik, Universit\"{a}t Bielefeld, Postfach 100131, 33501 Bielefeld, Germany\goodbreak
        \and
        GEPI \& ORN, Observatoire de Paris, Universit\'{e} PSL, CNRS, 5 Place Jules Janssen, 92190 Meudon, France\goodbreak
        \and
        Department of Physics \& Electronics, Rhodes University, PO Box 94, Grahamstown, 6140, South Africa\goodbreak
        \and 
        School for Physical Sciences, The Open University, Milton Keynes MK11 9DL, England\goodbreak
        \and
        Space Science and Technology Division, RALSpace, STFC Rutherford Appleton Laboratory, Chilton, Didcot OX11 0NX, England\goodbreak
        \and
        ORN, Observatoire de Paris, CNRS, PSL, Universit\'{e} d'Orl\'{e}ans, Nan\c{c}ay, France\goodbreak
        \and
        Ruhr University Bochum (RUB), Faculty of Physics and Astronomy, Astronomical Institute, Universit\"{a}tsstr. 150, 44801 Bochum, Germany\goodbreak
        \and
        Dipartimento di Fisica e Astronomia (DIFA), Universit\'{a} di Bologna, via Gobetti 93/2, 40129, Bologna, Italy\goodbreak 
        \and
        Istituto Nazionale di Astrofisica (INAF) - Istituto di Radioastronomia (IRA), via Gobetti 101, 40129, Bologna, Italy\goodbreak 
        \and
        Hamburger Sternwarte, Universit\"{a}t Hamburg, Gojenbergsweg 112, 21029, Hamburg, Germany\goodbreak
        \and
        Astronomical Observatory of the Jagiellonian University, ul. Orla 171, 30-244 Krak\'{o}w, Poland\goodbreak
        \and
        Th\"{u}ringer Landessternwarte, Sternwarte 5, D-07778 Tautenburg, Germany\goodbreak
        \and
        National Radio Astronomy Observatory, 520 Edgemont Road, Charlottesville, VA 22903, USA\goodbreak
        \and 
        CSIRO Space \& Astronomy, PO Box 1130, Bentley, WA 6102, Australia\goodbreak
        \and 
        Space Radio-Diagnostics Research Centre, University of Warmia and Mazury, ul.Oczapowskiego 2, 10-719 Olsztyn, Poland\goodbreak
        \and
        Institut f\"{u}r Theoretische Physik und Astrophysik, Universit\"{a}t W\"{u}rzburg, Emil-Fischer-Stra\ss e 31, 97074, W\"urzburg, Germany\goodbreak
        \and
        Institute of Astronomy, Faculty of Physics, Astronomy and Informatics, NCU, Grudziadzka 5, 87-100 Toru\'{n}, Poland\goodbreak
        \and 
        School of Physical Sciences, The Open University, Walton Hall, Milton Keynes, MK7 6AA, UK\goodbreak
        \and
        Centre for Extragalactic Astronomy, Department of Physics, Durham University, Durham DH1 3LE, UK\goodbreak 
        \and
        Institute for Computational Cosmology, Department of Physics, University of Durham, South Road, Durham DH1 3LE, UK\goodbreak
        \and
        Astronomy Department, Universidad de Concepci\'{o}n, Casilla 160-C, Concepci\'{o}n, Chile\goodbreak
        \and
        Instituto de Astrof\'{i}sica de Andaluc\'{i}a (IAA-CSIC), Glorieta de la Astronom\'{i}a s/n, E-18008 Granada, Spain\goodbreak
        \and
        Facultad de Ciencias, Universidad de Zaragoza, Pedro Cerbuna 12, E-50009 Zaragoza, Spain\goodbreak
        \and
        School of Sciences, European University Cyprus, Diogenes street, Engomi, 1516 Nicosia, Cyprus\goodbreak
        \and
        Astronomy and Astrophysics Section, Dublin Institute for Advanced Studies, 31 Fitzwilliam Place, Dublin D02 XF86, Ireland\goodbreak
        \and
        Anton Pannekoek Institute for Astronomy, University of Amsterdam, Science Park 904, 1098 XH Amsterdam, The Netherlands\goodbreak
        \and
        Inter-University Institute for Data Intensive Astronomy, Department of Astronomy, University of Cape Town, 7701 Rondebosch, Cape Town, South Africa\goodbreak
        \and
        Inter-University Institute for Data Intensive Astronomy, Department of Physics and Astronomy, University of the Western Cape, 7535 Bellville, Cape Town, South Africa\goodbreak
             }

   \date{Received 28 November 2022 / Accepted 17 December 2022}

\abstract{We present the detection of 68 sources from the most sensitive radio survey in circular polarisation conducted to date. We use the second data release of the 144\,MHz LOFAR Two-metre Sky Survey to produce circularly-polarised maps with median 140\,$\mu$Jy\,beam$^{-1}$ noise and resolution of 20$''$ for $\approx$27\% of the northern sky (5634\,deg$^{2}$). The leakage of total intensity into circular polarisation is measured to be $\approx$\,0.06\%, and our survey is complete at flux densities $\geq1$\,mJy. A detection is considered reliable when the circularly-polarised fraction exceeds 1\%. We find the population of circularly-polarised sources is composed of four distinct classes: stellar systems, pulsars, active galactic nuclei, and sources unidentified in the literature. The stellar systems can be further separated into chromospherically-active stars, M dwarfs, and brown dwarfs. Based on the circularly-polarised fraction and lack of an optical counterpart, we show it is possible to infer whether the unidentified sources are likely unknown pulsars or brown dwarfs. By the completion of this survey of the northern sky, we expect to detect 300$\pm$100 circularly-polarised sources.
}

\authorrunning{Callingham et al.}
\maketitle

%

\section{Introduction}

The electric field of radio emission received from distant sources can have a preferential orientation or rotation, which can be parameterised into total intensity (Stokes\,I), linear polarisation (Stokes\,Q\,and\,U), and circular polarisation (Stokes\,V). The polarised signal from radio sources can provide information about the emission mechanism, propagation effects, and physical properties of the radio sources not encoded in total intensity \citep{wielebinski2012}. 

Recently, there has been a particular resurgence of interest in detecting circularly-polarised radio sources. This renewed interest is largely because a high-degree of circular polarisation is an expected signpost of radio emission from extrasolar coronal-mass ejections \citep{osten17,2018ApJ...856...39C,2019ApJ...871..214V}, chromospherically-active stars \citep{Slee2003,2017ApJ...836L..30L,Zic2019}, pulsars \citep{1969Natur.221..724C,1998MNRAS.301..235G,1998MNRAS.300..373H}, and aurorae associated with the Solar System planets \citep{1998JGR...10320159Z,2009JGRA..114.8216F,2010JGRA..115.9221L}, exoplanets \citep{2001Ap&SS.277..293Z,2004ApJ...612..511L,2007P&SS...55..598Z,2018MNRAS.478.1763L,2021A&A...645A..59T}, brown dwarfs \citep{2008ApJ...684..644H,2016ApJ...818...24K,2017ApJ...846...75P,vedanthambrowndwarf}, and star-planet magnetic interactions \citep{2013A&A...552A.119S,2018ApJ...854...72T,2020NatAs.tmp...34V,2021A&A...645A..77P}. For these different classes of objects, the circularly-polarised fraction generally exceeds 10\% and is indicative of a coherent radio emission mechanism operating \citep{Vedantham2020}. In comparison, the circularly-polarised fraction is $\lesssim$1\% for radio-bright active galactic nuclei \citep[AGN;][]{saikia1988,osullivan2013}, intraday variable sources \citep{2000ApJ...545..798M}, and diffuse Galactic structures \citep{2017PhRvD..96d3021E}.

Despite the information that can be derived from circularly-polarised emission, wide-field ($\gtrsim1$\,sr) radio surveys have predominantly only reported total intensity measurements \citep[e.g.][]{Condon1998,Mauch2003,Murphy2010,gleam2017}. A wide-field, flux-density limited survey ensures an impartial sampling of the different classes of radio sources. Ignorance of which sources are observed is vital for maximising the detection rate of sources with beamed emission, such as pulsars and objects with radio aurorae. Since a priori we do not know which of these beamed systems are aligned with our line of sight, observing as many systems as possible maximises chance alignment. Furthermore, since the circularly-polarised radio sky is sparse compared to the total intensity sky, associations between dense optical surveys and radio sources can be reliably performed despite the low astrometric accuracy common to wide-field radio surveys \citep{2019RNAAS...3...37C}. 

Due to the increase in science opportunities and recent development of radio telescopes with large fields of view, there has been new momentum in producing wide-field circularly-polarised radio surveys. The Australian Square Kilometre Array Pathfinder (ASKAP) telescope has conducted the Rapid ASKAP Continuum Survey \citep[RACS;][]{2020PASA...37...48M}, a survey covering the sky south of declination $+41^{\circ}$ at $\approx$900\,MHz with a sensitivity of $\approx$250 $\mu$Jy\,beam$^{-1}$. Preliminary analysis of the Stokes\,V maps, only reporting significant circularly-polarised sources that also had an optical counterpart, predominantly identified magnetically-active M\,dwarfs, pulsars, close binaries, and chemically-peculiar A- and B-type
stars \citep{pritchard2021}. Such a population of sources is consistent with the type of radio-bright stellar systems identified at 1.4\,GHz \citep{1999AJ....117.1568H}.

However, if one of the scientific drivers of a wide-field survey is to identify radio aurorae associated with star-planet interactions and exoplanets, conducting a survey at low frequencies offers several advantages \citep{2011RaSc...46.0F09G,2011A&A...531A..29H}. Firstly, the electron-cyclotron maser instability (ECMI) mechanism responsible for the auroral radio emission has a high-frequency spectral cut-off that is directly proportional to the magnetic-field strength of the emitting body \citep{2006A&ARv..13..229T}. For example, the highly-circularly polarised radio emission produced by the Jupiter-Io interaction is not observable above 40\,MHz because the Jovian surface magnetic field strength at the Io magnetic footprint does not exceed $\approx$14\,G \citep{1983phjm.book..226C,marques2017}.

Secondly, conducting observations at low frequencies can aid in distinguishing between coherent emission mechanisms. Two of the main mechanisms that produce coherent radio emission from stellar systems are the ECMI and plasma emission \citep{1985ARA&A..23..169D}. While both emission mechanisms produce radio emission with a high degree of circular polarisation, fundamental plasma emission is limited to brightness temperatures of $\lesssim$10$^{12}$\,K for even the most chromospherically active stars \citep{1985ARA&A..23..169D,Stepanov2001,Vedantham2020}. Since brightness temperature $T_{B}$ $\propto$ $\nu^{-2}$, where $\nu$ is frequency, low-frequency observations are more powerful than gigahertz-frequency observations for differentiating between emission mechanisms at a given sensitivity \citep[e.g.][]{2020NatAs.tmp...34V}. 

While conducting a low-frequency circularly-polarised survey is advantageous for stellar system science, the relatively low sensitivity and resolution of low-frequency arrays has largely prevented such surveys from being performed. The only previous low-frequency wide-field circularly-polarised survey was conducted by the Murchison Widefield Array \citep[MWA;][]{Tingay2013}, covering the sky south of declination $30^{\circ}$ at 200\,MHz \citep{2018MNRAS.478.2835L}. However, due to the survey's low sensitivity of $\approx$3\,mJy\,beam$^{-1}$, \citet{2018MNRAS.478.2835L} only detected previously known pulsars, artificial satellites, and Jupiter. The non-detection of stellar systems by \citet{2018MNRAS.478.2835L} is consistent with the expectation that the flux densities of radio aurorae from star-planet interactions and exoplanets at $\approx200$\,MHz are likely $\lesssim$1\,mJy \citep{2011RaSc...46.0F09G,2012MNRAS.427L..75N,2018MNRAS.478.1763L,2019MNRAS.484..648P}.  

With the LOw-Frequency ARray \citep[LOFAR;][]{vanHaarlem2013}, we are currently conducting the 144\,MHz LOFAR Two-metre Sky Survey \citep[LoTSS;][]{2016arXiv161102700S,2019A&A...622A...1S}. LoTSS will cover the entire northern sky with an unprecedented root-mean-square noise of $\approx$100\,$\mu$Jy\,beam$^{-1}$ in full-Stokes polarisation. The unique discovery space that LoTSS provides has already been exemplified by the several studies already published using preliminary Stokes\,V data, such as the first direct detection of a brown dwarf from its radio properties \citep{vedanthambrowndwarf}, the inference of a star-planet interaction \citep{2020NatAs.tmp...34V,2022MNRAS.514..675K}, long-term auroral emission from an active M dwarf system \citep{2021A&A...648A..13C}, evidence of closed-field lines extending to large stellar radii around an M dwarf \citep{2021A&A...650L..20D}, new pulsars missed in previous time-domain searches \citep{2022A&A...661A..87S}, and coherent emission detected from chromospherically-active binaries \citep{toet21}. 

In this paper we present V-LoTSS, the circularly-polarised component of LoTSS, using the second LoTSS data release \citep[DR2;][]{lotss-dr2}. In Section\,\ref{sec:data_red}, we summarise the observations and data reduction strategy, including the leakage analysis. We then detail the source finding and quality assurance steps performed to produce the V-LoTSS catalogue in Section\,\ref{sec:cat}. Section\,\ref{sec:discuss} features our discussion about the astrophysical composition of V-LoTSS, the ability to discern the nature of unidentified circularly-polarised sources, and the source counts of V-LoTSS. Finally, we conclude in Section\,\ref{sec:conc}, with an additional outline of the data processing improvements we can perform for future V-LoTSS releases.

\section{Observations and data reduction}
\label{sec:data_red}

The sky coverage of this release of V-LoTSS is identical to that of the second data release (DR2) of LoTSS \citep{lotss-dr2}. LoTSS-DR2 covers 5634 square degrees, which is composed of two contiguous regions of sky centred at approximately 12h45m00s +44$^\circ$30$\arcmin$00$\arcsec$ and 1h00m00s +28$^\circ$00$\arcmin$00$\arcsec$. The LoTSS-DR2 sky coverage is composed of 841 individual $\approx$8\,h LOFAR pointings. All of these pointings are at Galactic latitudes $>$20$^{\circ}$.

The calibration and data reduction routines followed for V-LoTSS are identical to those outlined by \citet{2019A&A...622A...1S} and \citet{lotss-dr2}, and was performed during the processing of LoTSS-DR2. In summary, all observations had a 1\,s integration time and 12.1875\,kHz spectral resolution covering 120 to 168\,MHz. After the standard direction independent (DI) calibration pipeline\footnote{\url{https://github.com/lofar-astron/prefactor}}, the data are averaged to a frequency resolution of 0.195\,MHz and a time resolution of 8\,s to reduce the data volume. Since we are only producing 6$''$ or lower resolution images, this averaging is a compromise between minimising the impact of smearing and our processing capabilities. The DI calibration pipeline includes correcting any clock offsets between stations, phases, bandpass, and ionospheric Faraday rotation. Outside of the application of the primary beam, all polarisation calibration is performed during the direction-independent stage of processing.

Once the DI products are produced, a direction-dependent (DD) calibration and imaging pipeline\footnote{\url{https://github.com/mhardcastle/ddf-pipeline}} is employed to reach better sensitivity and dynamic range. The LoTSS-DR2 DD pipeline and its performance are described in detail by \cite{2021A&A...648A...1T}, employing DDFacet \citep{Tasse2018} and killMS \citep{Tasse2014}. In summary, a sky model is produced from a DI image of the 8.3$^{\circ}$ $\times$ 8.3$^{\circ}$ wide field, which is used to simultaneously produce calibration solutions in 45 directions. The field is then imaged again after applying these newly derived phase corrections. An intermediate step is done to ensure the flux density scale is set to the \citet{Scaife2012} flux density scale \citep{Hardcastle2016}. Amplitude and phase corrections are then applied after another round of derivation of DD calibration solutions using this updated sky model. This involves applying the primary beam model and astrometric corrections for any per-facet level ionospheric effects via crossmatching to the Pan-STARRS optical catalogue \citep{lotss-dr2}.

20$''$ resolution full-bandwidth Stokes\,I and V images are made using DDFacet. Deconvolved Stokes\,I images at $6''$ resolution are also produced, from which a catalogue of 4,396,228 $\geq$5$\sigma$ sources was formed after mosaicing the individual fields, where $\sigma$ represents the rms noise. The Stokes\,I data of LoTSS-DR2 has an astrometric accuracy of $\approx$0.2$''$, an rms noise better than 171\,$\mu$Jy in 95\% of its area, is 95\% complete at $\approx$1\,mJy, and has a flux density scale accurate to within 10\% \citep{lotss-dr2}. 

At the time of producing this release of V-LoTSS, DDFacet did not have the capabilities to deconvolve the Stokes\,V data and we only produced Stokes\,V images at 20$''$ resolution \citep[see][for details]{Tasse2018}. Since our science goals for conducting this survey are focused on detecting unresolved circularly-polarised sources, and the sky density of such objects are so low, not deconvolving the Stokes\,V data does not severely limit the science applications of the images. However, future releases of V-LoTSS will be based on deconvolved Stokes\,V images, facilitating such secondary science goals as searching for circular polarisation from hotspots embedded in the resolved lobes of radio-loud AGN. 

\subsection{Polarisation leakage}
\label{sec:leakage}
Antenna imperfections, poor primary beam models, and path effects can result in leakage of total intensity into the different polarisation vectors. Since LOFAR is composed of linear dipoles, the leakage into Stokes\,V due to mechanical flaws is expected to be small. The polarisation response of the primary beam of a low-frequency aperture array is also notoriously difficult to model and measure since the primary beam is sky variable, has relatively large sidelobes, and has significant antenna cross-coupling \citep{2016ApJ...826..199N,2017PASA...34...62S}. 

To test the amount of leakage into Stokes\,V, we measured the absolute peak pixel in a 20$''$ radius circular region at the location of every unresolved and isolated $\geqslant 5\sigma$ Stokes\,I source \citep[according to Eqn. 1 derived by][]{lotss-dr2} in the corresponding primary-beam corrected Stokes\,V image. The radius was set to 20$''$ due to the resolution of the Stokes\,V images, ensuring that we captured any leakage potentially offset from the Stokes\,I source. We then took the ratio of the flux density of the absolute peak Stokes\,V pixel $S_{V,p}$ to the total flux density of the Stokes\,I source $S_{I}$, where $S_{I}$ is measured in each individual LoTSS field. The total flux density was used for Stokes\,I since we wanted to account for any small ionospheric blurring produced from the higher-resolution measurement.

\begin{figure}
\begin{center}
\includegraphics[scale=0.25]{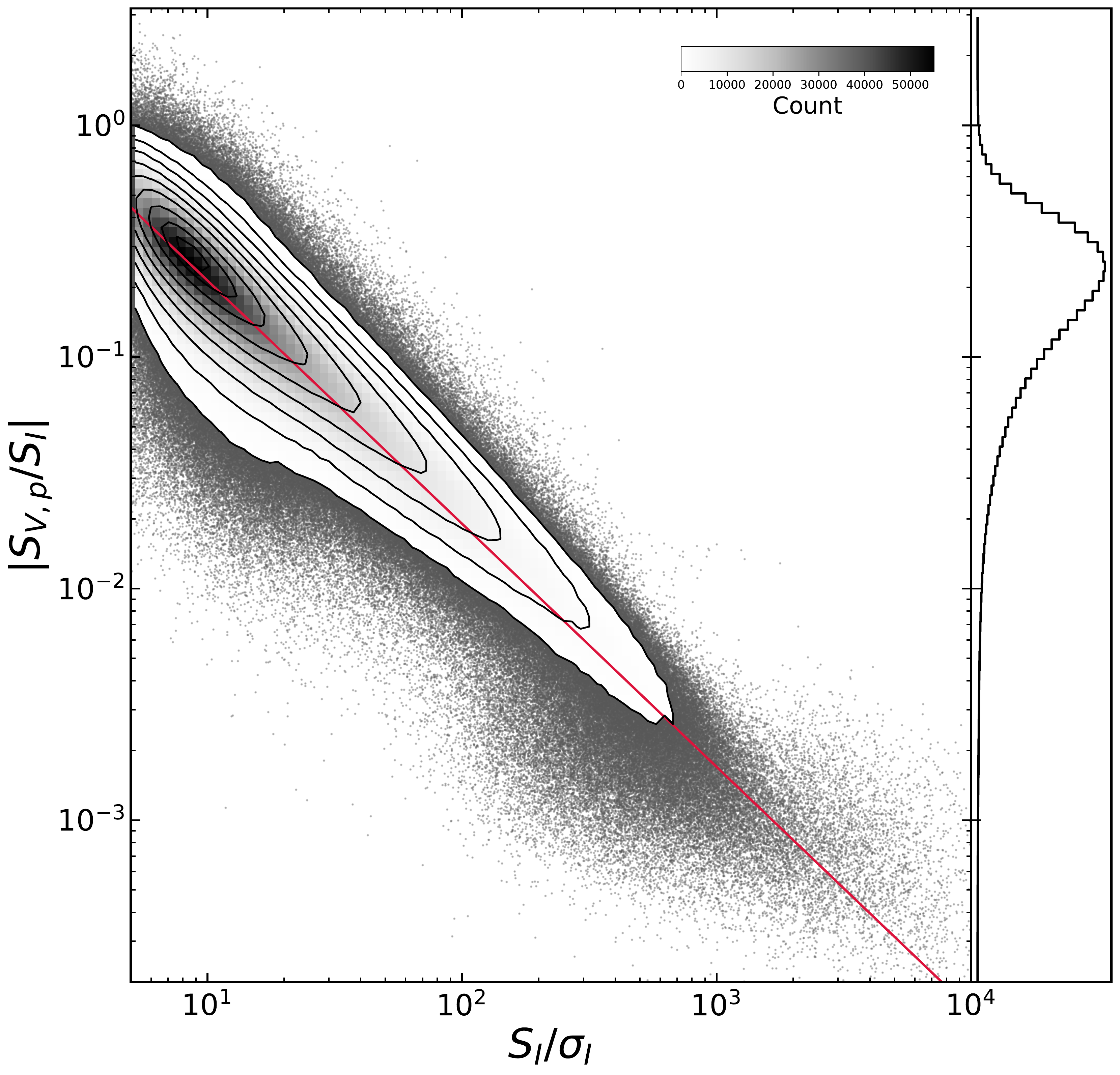}
 \caption{The absolute ratio of the maximum flux density of a Stokes\,V pixel $S_{V,p}$ to the total flux density of the corresponding Stokes\,I source $S_{I}$ as a function of signal-to-noise ratio $S_{I}/\sigma_{I}$ in Stokes\,I. The contours associated with the density plot correspond to 300, 900, 2500, 5000, 10000, 20000, 30000, 40000, and 50000 measurements. The colour bar at the top right-hand corner indicates the number of counts in the density plot. The red line is the trend measured when the aperture used to calculate the maximum Stokes\,V signal is replaced with Gaussian random noise. The simulation follows the measured data well up to $S_{I}/\sigma_{I} \sim 10^{3}$, the point at which leakage becomes dominant. The histogram in the right panel shows the distribution of $|S_{V,p}/S_{I}|$, which is dominated by low signal-to-noise sources.}
\label{fig:leakage_snr}
\end{center}
\end{figure}

In Figure\,\ref{fig:leakage_snr} we plot the ratio of the flux density of the maximum Stokes\,V pixel to a corresponding Stokes\,I source flux density as a function of Stokes\,I signal-to-noise ratio. As expected, noise bias dominates at $\lesssim 20 S_{I}/\sigma_{I}$, where $\sigma_{I}$ is the local rms noise in Stokes\,I. In this low signal-to-noise regime, the maximum pixel in Stokes\,V corresponds to a noise peak in a finite aperture, while the Stokes\,I flux density corresponds to a real source. The lower the signal-to-noise of the source, the closer its flux density will be to this noise peak. We confirmed this bias was present for sources with signal-to-noise ratios $\lesssim 10^{3}$ by simulating random Gaussian noise in identically-sized apertures in Stokes\,V and repeating the measurement steps above. The trend present in these simulations is presented as the red line in Figure\,\ref{fig:leakage_snr}. The offset to larger values of $|S_{V,p}/S_{I}|$ at Stokes\,I signal-to-noise ratios greater than $10^{3}$ implies this is a direct measurement of the leakage. 

If no leakage was present, the relationship between $|S_{V,p}/S_{I}|$ and $\sigma_{I}/ S_{I}$ would only be influenced by noise. Instead, we find the relation:

\begin{equation}
    \log\left(|S_{V,p}/S_{I}|\right) = A\log\left(\sigma_{I}/ S_{I}\right) + B,
    \label{eqn:leakage}
\end{equation}

\noindent where $A$ and $B$ encode the amount of noise bias and leakage, respectively. For the best fit of Equation\,\ref{eqn:leakage}, we measure $A = 1.8 \pm 0.1$ and $B = 0.06 \pm 0.03 \%$. To be conservative, we deem any source with a circularly-polarised fraction $\geqslant$1\% as reliable. The values for $A$ and $B$ are larger than those reported by \citet{lotss-dr2} because we measured the leakage in lower Stokes\,I signal-to-noise sources, used total flux density for the Stokes\,I measurement, and searched to larger angular distances from the pointing centre.

\begin{figure}
\begin{center}
\includegraphics[scale=0.25]{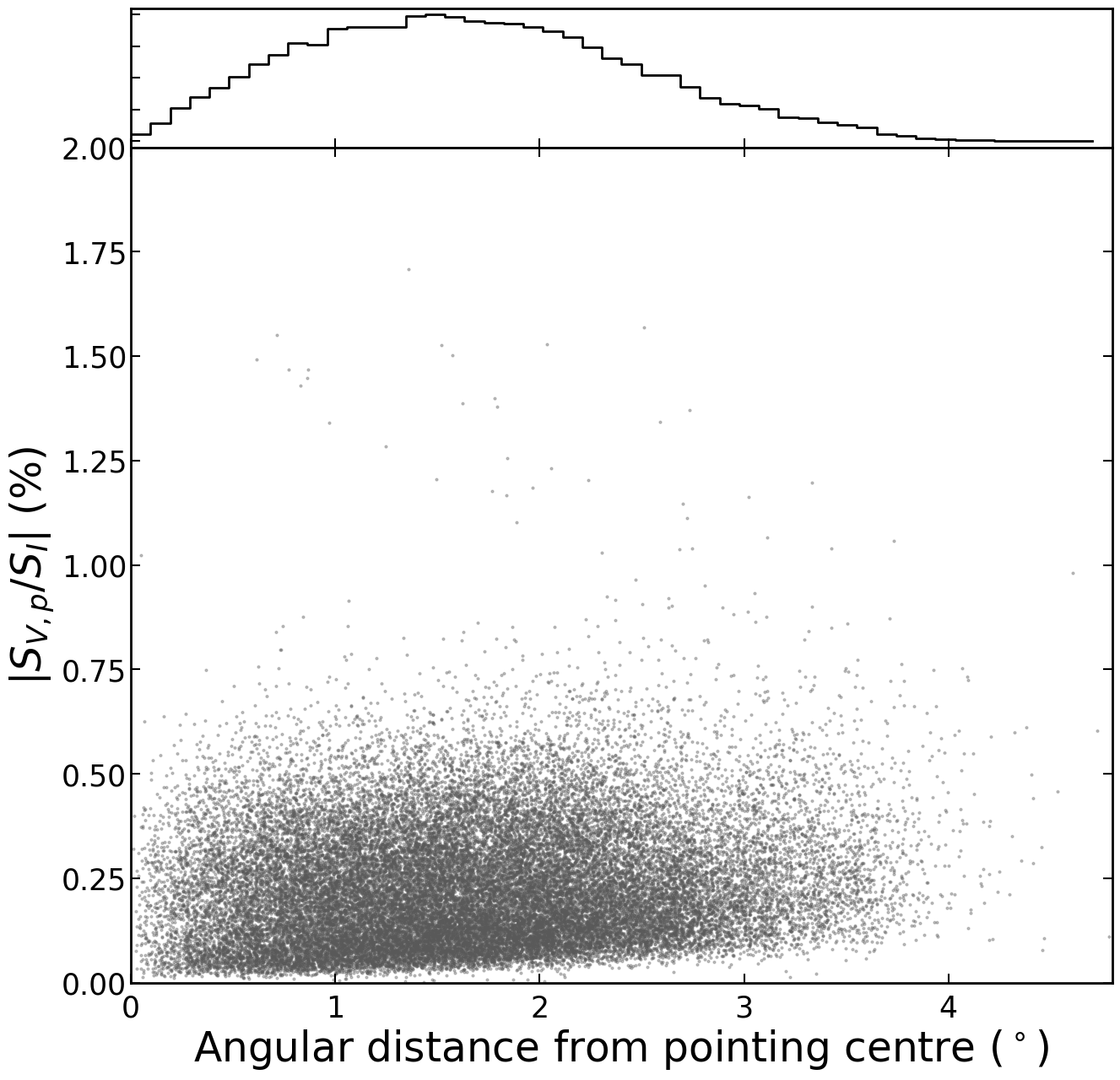}
 \caption{The absolute ratio of the maximum flux density of a Stokes\,V pixel $S_{V,p}$ to the total flux density of a corresponding Stokes\,I source $S_{I}$ as a function of angular distance from the LoTSS field pointing centre for sources with Stokes\,I signal-to-noise ratio $\geqslant 500$. The top histogram shows the relative number of measurements with angular distance. We find a weak dependence of leakage with angular distance out to 3$^{\circ}$. After $\approx$3$^{\circ}$ angular distance, non-Gaussian noise and poorly calibrated facets account for the dependence.} 
\label{fig:leakage_pointing}
\end{center}
\end{figure}

Since the primary beam model can impact the amount of leakage, we also show the distribution of $|S_{V,p}/S_{I}|$ against angular distance from the pointing centre of a LoTSS field in Figure\,\ref{fig:leakage_pointing}. We observe a weak dependence of leakage with distance from the LoTSS pointing centre out to $\approx$3$^{\circ}$, corresponding to the $\approx$25\% power point of the primary beam. At distances greater than 3$^{\circ}$, and especially at distances greater than 4$^{\circ}$, non-Gaussian noise is present since distant facets lack enough high signal-to-noise sources for accurate calibration. Therefore, we restrict our search for significant Stokes\,V sources to angular distances less than 3$^{\circ}$ from the pointing centre of a LoTSS field. 

We perform our Stokes\,V search on individual LoTSS fields, as opposed to an overall mosaic of the fields, because the emission of many of our sources of interest is time variable. In making a mosaic, where the individual fields are averaged in the image plane, it is possible to wash out a detection since the source may not be emitting in Stokes\,V in an adjoining field that was observed on a different date.

\begin{figure*}
\begin{center}$
\begin{array}{c}
\includegraphics[scale=0.4]{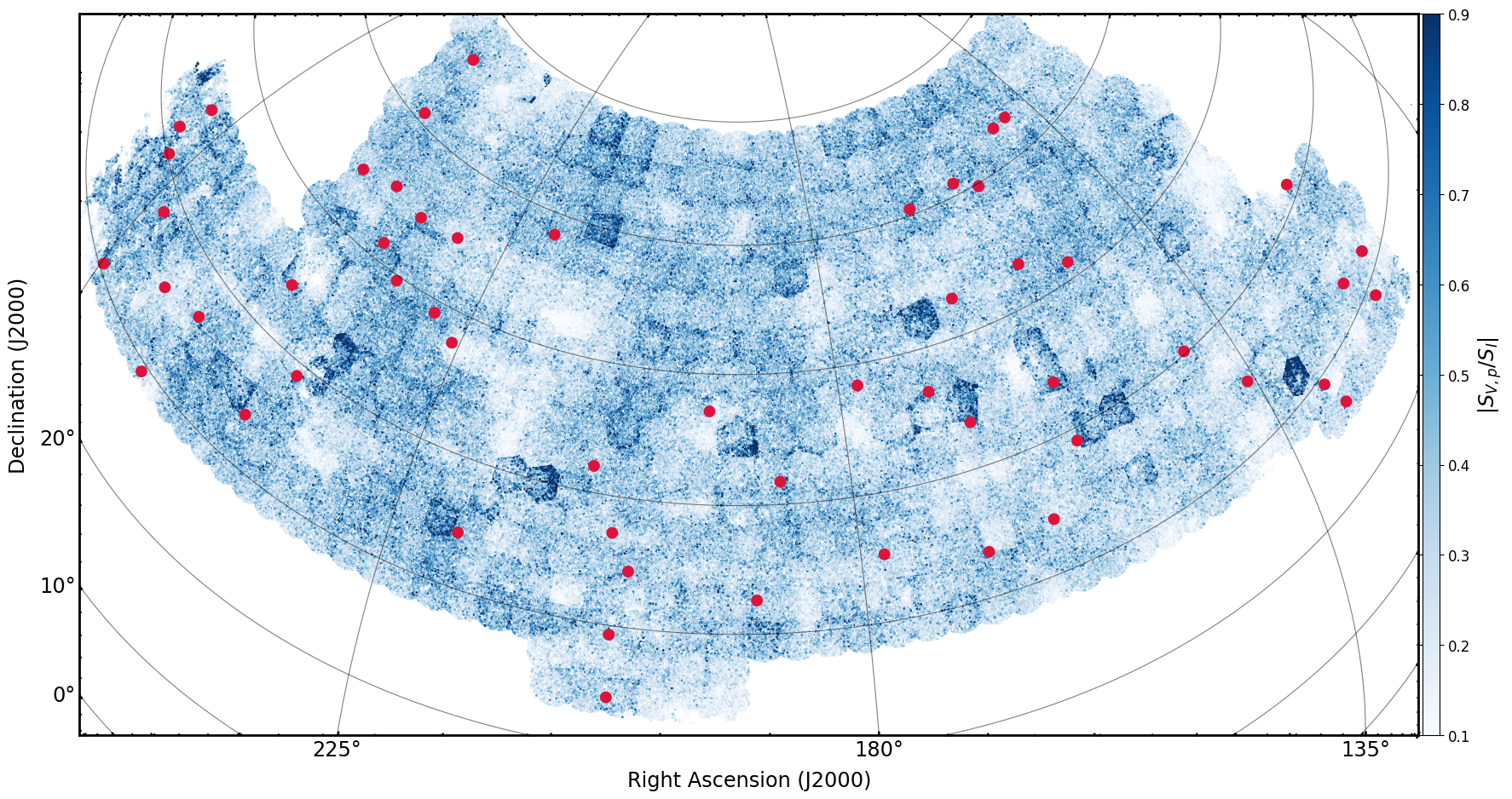}\\
\includegraphics[scale=0.373]{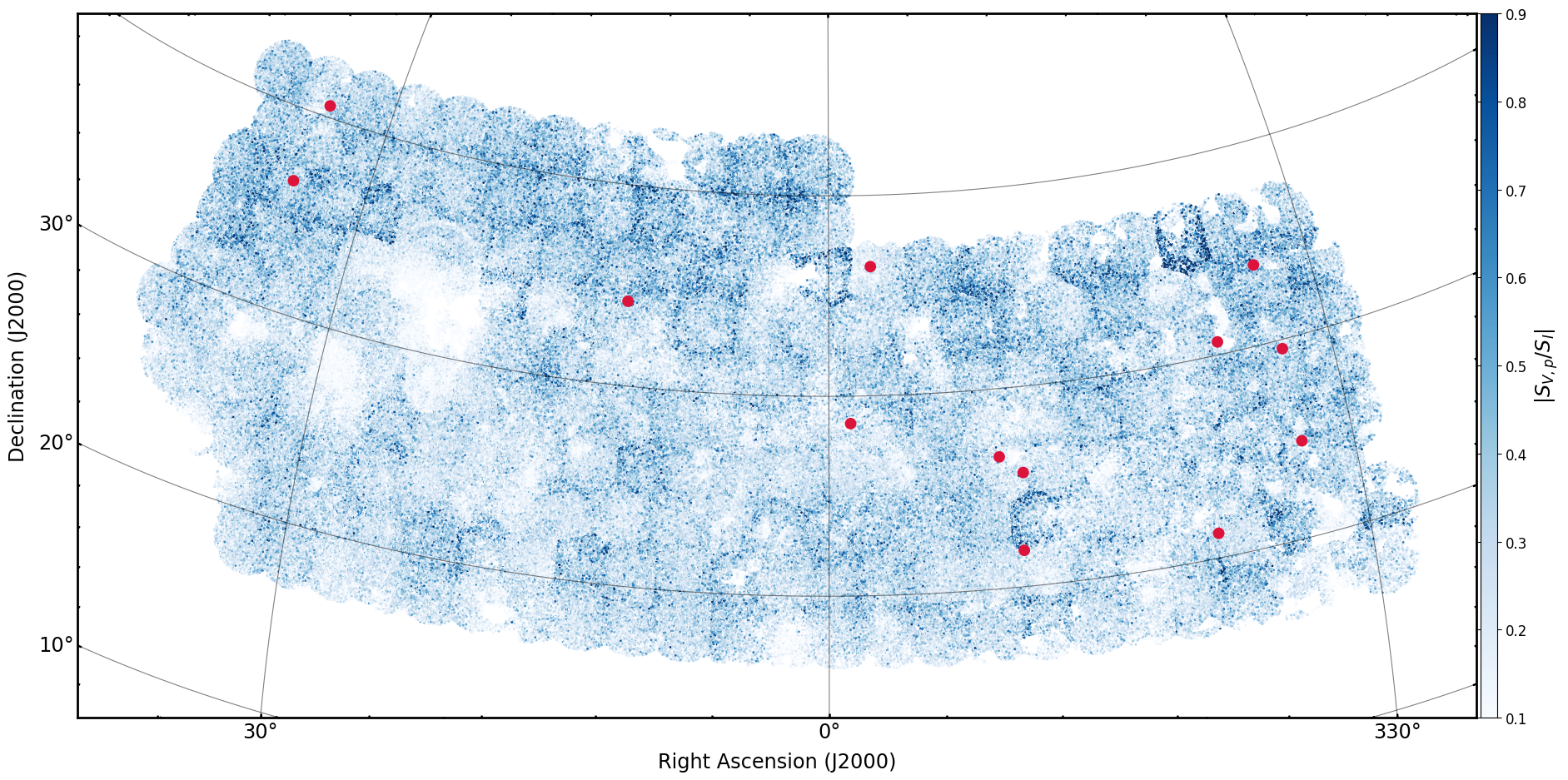}\\
\end{array}$
 \caption{The sky distribution of $|S_{V,p}/S_{I}|$ for the fields centred on RA 13\,h (top panel) and RA 0\,h (bottom panel). The significant Stokes\,V sources we detect are shown as red circles.}
\label{fig:leakage_radec}
\end{center}
\end{figure*}

Finally, we can identify any individual problematic fields by considering the sky distribution of our leakage measurements. We show in Figure\,\ref{fig:leakage_radec} the sky distribution of $|S_{V,p}/S_{I}|$, along with the positions of our detected Stokes\,V sources discussed in Section\,\ref{sec:cat}. While we expect some localised correlation per-field, we remove any outlier fields from our Stokes\,V source finding that have a median leakage that exceeds 1\%. That was the case for only one field: P150+40. This field was observed during poor ionospheric conditions and is thus excluded from the source finding detailed in Section\,\ref{sec:cat}.  

\subsection{Noise Properties of V-LoTSS}

The Stokes\,I component of LoTSS-DR2 has a sensitivity better than $\approx$100\,$\mu$Jy over $\approx$70\% of its area \citep{lotss-dr2}. Our Stokes\,V images should have noise closer to the thermal limit of the observations since the images will not have the sidelobe confusion or deconvolution artefacts that are present in the Stokes\,I images. However, we are searching for significant Stokes\,V sources on a per-field basis. This means that we do not gain the benefit of reducing the noise when forming a mosaic. 

\begin{figure}
\begin{center}
\includegraphics[scale=0.32]{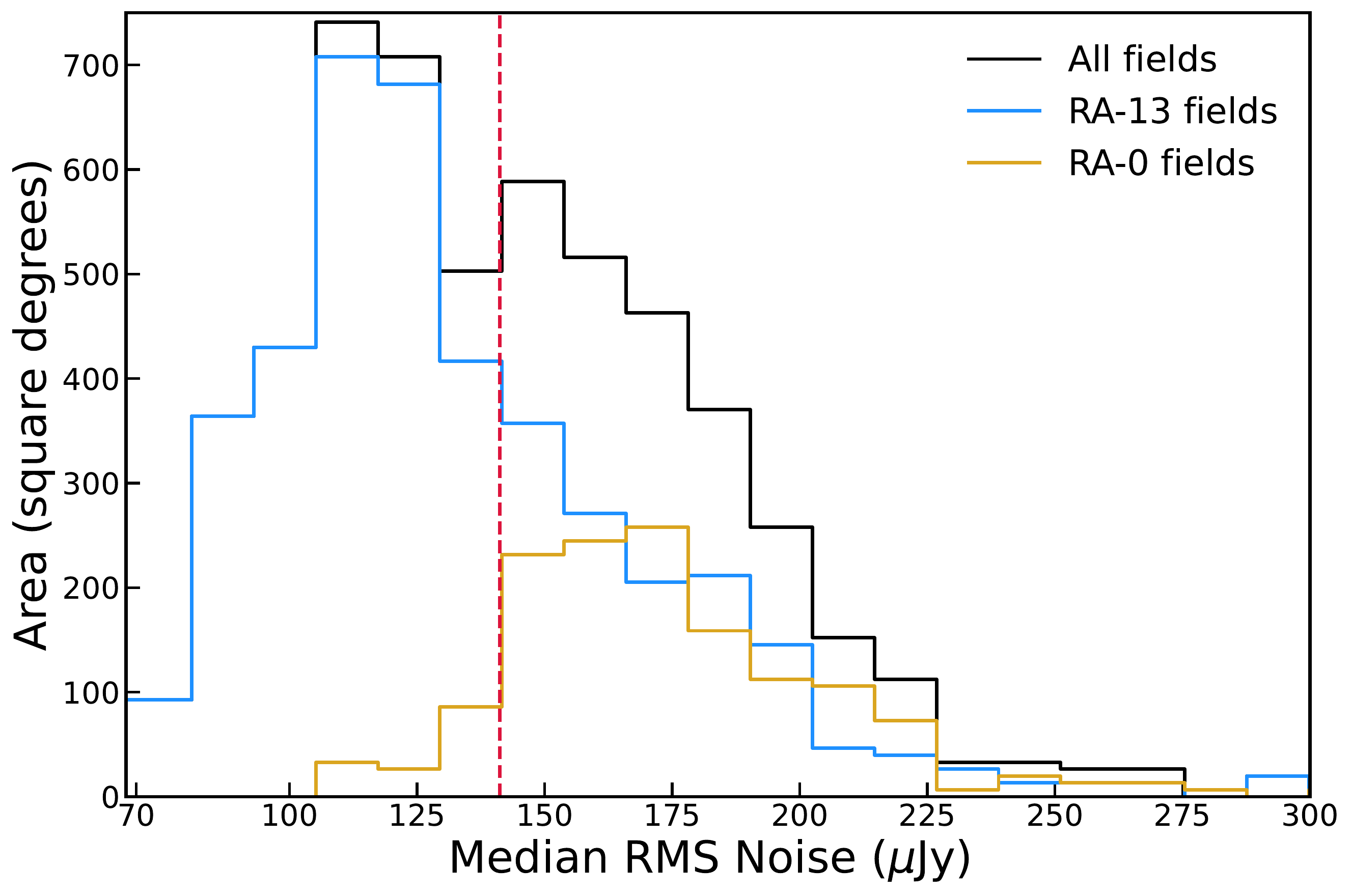}
 \caption{The distribution of the median rms noise within 3$^{\circ}$ of the pointing centre for the primary-beam corrected V-LoTSS images. The group of fields centred around an RA of 13h are shown in blue, while the fields centred around an RA of 0h are shown in yellow. The distribution of all fields is shown in black. The median rms noise for all the fields is plotted as a red-dashed line at 140\,$\mu$Jy\,beam$^{-1}$.} 
\label{fig:rms}
\end{center}
\end{figure}

We believe that a fair reflection of the rms noise present in our Stokes\,V images is the median noise within 3$^{\circ}$ of the pointing centre of the field. We show the distribution of the V-LoTSS rms noise in Figure\,\ref{fig:rms}. We find that the median rms for all fields to be 140$^{+32}_{-27}$\, $\mu$Jy\,beam$^{-1}$, where the uncertainties are the 75th and 25th percentile of the distribution. 95\% of all the Stokes\,V fields have a median rms noise better than 220\,$\mu$Jy\,beam$^{-1}$. As expected, the fields centred around a RA of 0h have a higher median rms noise than the fields centred around a RA of 13h. This is because the RA 0h fields were observed at a lower elevation relative to the RA 13h fields.

For a direct field-to-field comparison between the Stokes\,I and V components, we also repeated the above analysis but only within 0.2$^{\circ}$ of the LoTSS pointing centre. The small angular distance of 0.2$^{\circ}$ was chosen since the noise in an image would not be significantly improved by mosaicing at such a distance. We find 95\% of the rms noise of the Stokes\,V images to be better than 150\,$\mu$Jy\,beam$^{-1}$, a 21\,$\mu$Jy\,beam$^{-1}$ improvement over the Stokes\,I image rms noise in the same area.

\subsection{Astrometry of V-LoTSS}
\label{sec:astrometry}

Since the Stokes\,V products are derived from the astrometrically-corrected Stokes\,I products, the absolute astrometric accuracy is identical. The astrometric accuracy of LoTSS-DR2 is $\approx$ 0.2$''$ \citep{lotss-dr2}. However, the Stokes\,V images have a lower resolution of 20$''$, compared to the 6$''$ resolution of the Stokes\,I images. Therefore, the positional uncertainty of a Stokes\,V source will be nearly four times larger than its Stokes\,I counterpart if the signal-to-noise of the source is identical in Stokes\,V and Stokes\,I. 

We deem a match out to 4$''$ between a Stokes\,V and I source to be reliable down to a signal-to-noise ratio of 4 in Stokes\,V following the formulation of \citet{Condon1998}. 

\subsection{Flux Density Scale of V-LoTSS}

The absolute flux density scale of V-LoTSS is identical to that of LoTSS since we scaled the images of V-LoTSS and corresponding individual Stokes\,I fields using the scalings derived by \citet{lotss-dr2}. \citet{lotss-dr2} found the flux density scale to be better than 10\%. Since we find no sky-dependence of leakage, as shown in Figure\,\ref{fig:leakage_radec}, we infer we also have an accurate model of the Stokes\,V primary beam. Therefore, we consider the flux density scale of V-LoTSS to also be better than 10\%.  

Note that since we have so few Stokes\,V detections, which are likely time variable, an independent test of the absolute flux density scale in Stokes\,V would not be conclusive.

\subsection{Definition of the handedness of Stokes V emission}

We define the sign of circularly-polarised light as left-hand circularly-polarised light minus right-hand circularly-polarised light. Since there are several situations in the production of V-LoTSS where a sign-convention decision can influence the final handedness, we ensured that our Stokes\,V measurements are consistent with our definition by checking the observed polarity of emission from known circularly-polarised pulsars \citep{1998MNRAS.301..235G,1998MNRAS.300..373H,2010PASA...27..104V,2015A&A...576A..62N}. The pulsars in our survey footprint that we detect from these studies (see Section\,\ref{sec:psr_cat} for details) are B0045+33, B0751+32, B0823+26, B0917+63, J1012+5307, B2210+29, and B2315+21. We excluded B0751+32 and B2315+21 from our analysis as their circularly-polarised handedness reverses between two of the references \citep{1998MNRAS.301..235G,1998MNRAS.300..373H}. For the remaining pulsars, we measured the same handedness of the circularly polarised light as that reported in the literature \citep{1998MNRAS.301..235G,1998MNRAS.300..373H,2010PASA...27..104V,2015A&A...576A..62N}, implying that our Stokes\,V sign convention is consistent with left-hand circularly-polarised light minus right-hand circularly-polarised light.
  
\section{The V-LoTSS catalogue}
\label{sec:cat}
\subsection{Source finding and quality assurance}
\label{sec:source_finding_QA}

We searched for significant sources in the Stokes\,V images using the Background And Noise Estimator (\textsc{BANE}) and source finder \textsc{Aegean} \citep[v\,2.2.3;][]{2012MNRAS.422.1812H,2018PASA...35...11H}. We seeded \textsc{Aegean} with a 4$\sigma_{V}$ clip, where $\sigma_{V}$ is the local rms noise in the Stokes\,V image, and a symmetric synthesised beam of 20$''$ full-width half-maximum. A 5$\sigma_{V}$ flood clip was applied.

In each Stokes\,V field, we searched for significant sources in a circular region with a 3$^{\circ}$ radius. This area corresponds to searching over 18.1 million pixels per field. Therefore, we expect at least eleven $\geqslant 5\sigma_{V}$ noise spikes per field, and at least $\approx$9,240 $\geqslant 5\sigma_{V}$ noise spikes over our whole survey area. We find a total of 20,118 $\geqslant 5\sigma_{V}$ detections, which is over double the prediction due to the presence of real sources, leakage, and non-Gaussian noise present around bright Stokes\,I sources.

To form a reliable sample of Stokes\,V detections we performed the following quality assurance steps:

\begin{enumerate}
    \item We ensured every Stokes\,V detection was associated with an unresolved $\geqslant 5\sigma_{I}$ Stokes\,I source. Requiring such an association increases the reliability of the Stokes\,V detection because the noise properties of the Stokes\,I and V images are independent when there is no leakage. As detailed in Section\,\ref{sec:astrometry}, we deemed a match to be true if the Stokes\,I position was within 4$''$ of the Stokes\,V position. With this 4$''$ matching radius, we expect $\lesssim1$ matches of a $\geqslant 5\sigma_{I}$ unresolved source from the 6$''$ resolution images with a $\geqslant 5\sigma_{V}$ noise spike over the entire survey region. We empirically confirmed that false association rate by applying random 10-20$'$ shifts to the Stokes\,V sources and repeating the crossmatch. By requiring our $\geqslant 5\sigma_{V}$ Stokes\,V detections to be associated with an unresolved $\geqslant 5\sigma_{I}$ Stokes\,I source, we are left with a sample of 3,008 sources.
    
    \item We removed any Stokes\,V source that was located further than 4$''$ but within 2$'$ of a Stokes\,I source that had a signal-to-noise ratio greater than 500. This signal-to-noise ratio corresponds roughly to a median total intensity flux density of 125\,mJy. This step is necessary because deconvolution errors around the brightest sources produce significant non-Gaussian rms noise and spurious Stokes\,I sources. The 2$'$ radius was derived heuristically after determining when the noise properties around bright Stokes\,I sources returned to Gaussian. This cut removed 1,183 sources, leaving 1,824 $\geqslant 5\sigma_{V}$ Stokes\,V detections. We note that this cut also reduces our survey area by $\approx$61 square degrees, or $\approx$0.2\%.
    
    \item We guaranteed that the circularly-polarised fraction of a source was not consistent with leakage by requiring $|S_{V}/S_{I}| \geqslant 1\%$, as detailed in Section\,\ref{sec:leakage}. This step left us with 102 $\geqslant 5\sigma_{V}$ Stokes\,V detections, of which 56 are unique as some sources were detected multiple times in neighbouring fields. Almost all the detections show variability between neighbouring fields.
\end{enumerate}

\subsection{Targeted association and source identification}
\label{sec:targ_assoc}

For sources with $4 \leqslant S_{V}/\sigma_{V} < 5$, it is difficult to form a reliable blind sample. This is because we expect $\sim$200 chance co-incidence matches between Stokes\,V noise peaks of $4 \leqslant S_{V}/\sigma_{V} < 5$ and unresolved Stokes\,I sources with flux density $\geqslant 5\sigma_{I}$. Following the same quality assurance steps outlined in Section\,\ref{sec:source_finding_QA}, we find 1,448 sources that are associated with a $\geqslant 5\sigma_{I}$ unresolved Stokes\,I source and have a circularly-polarised fraction $\geqslant 1\%$. 

However, it is possible to identify the reliable detections from these lower signal-to-noise sources by making assumptions about the nature of the sources. For example, we know that stars, ultracool dwarfs, and pulsars can emit circularly-polarised radio emission. By crossmatching these lower signal-to-noise sources to sources in the catalogues containing information about the location of a star, pulsar, or brown dwarf, we can identify the reliable detections. Furthermore, crossmatching our blind $\geqslant 5\sigma_{V}$ sample to these catalogues aids in source identification.

\subsubsection{\emph{Gaia} DR3 catalogue}

To identify any stars in our sample, we crossmatched all $\geqslant 4\sigma_{V}$ 1,504 V-LoTSS sources to the \emph{Gaia} Data Release 3 \citep[DR3;][]{gaiadr3} catalogue. We considered a match to be reliable if the proper-motion corrected position of a Galactic \emph{Gaia} DR3 source was less than 1.3$''$ from the Stokes\,I V-LoTSS source position. A \emph{Gaia} DR3 source was determined to be a Galactic object if it had a parallax over error of $\varpi / \varpi_{\mathrm{err}} \geqslant 3$ \citep{CallinghamPopulation}. We find 38 matches, which are mostly composed of M dwarfs, chromospherically-active binaries, and millisecond pulsars. The properties of this sample are explored in Section\,\ref{sec:stars}. Of these 38 matches, 29 are V-LoTSS sources with $\geqslant 5\sigma_{V}$. Only one source is a millisecond pulsar.

Since all the fields of V-LoTSS are located at Galactic latitudes $|b| > 20^{\circ}$, the average density of \emph{Gaia} DR3 sources with $\varpi / \varpi_{\mathrm{err}} \geqslant 3$ is $\approx$\,1300 per square~degree. Therefore, we would expect the number of chance-coincidence associations between \emph{Gaia} DR3 Galactic sources and the V-LoTSS sample to be $<0.8$. As performed by \citet{2019RNAAS...3...37C}, we also empirically confirmed this chance-coincidence rate by applying random 10$'$--20$'$ shifts to the \emph{Gaia} DR3 positions and performing the crossmatch again. Furthermore, all of the sources presented in this paper were also observed to display variability inconsistent with noise or proper motion shifts in neighbouring LoTSS fields \citep[see e.g.][]{CallinghamPopulation,toet21}. We consider all presented detections as reliable.

\subsubsection{Pulsar catalogues}
\label{sec:psr_cat}
To search for any pulsars in V-LoTSS, we crossmatched our detections to the following pulsar catalogues: the Australia Telescope National Facility Pulsar Catalogue \citep{2005AJ....129.1993M}, the LOFAR tied-array all-sky survey \citep[LOTAAS;][]{2018ApJ...866...54T,2020MNRAS.492.5878T,2020MNRAS.491..725M}, and the TULIPP survey \citep{2022A&A...661A..87S}. The compilation of these three catalogues results in 3177 unique pulsars, of which 98 are in the V-LoTSS survey footprint. Note that several of the new pulsar discoveries in the TULIPP survey were made using a preliminary version of the V-LoTSS catalogue.

We crossmatched out to 5$''$ and found 24 matches to known pulsars in V-LoTSS, implying a detection rate of $\approx$25\%. We would expect $<0.01$ of these matches to be a chance coincidence. Only three of these pulsar matches are to V-LoTSS sources with $4 \leqslant S_{V}/\sigma_{V} < 5$. This V-LoTSS pulsar sample is explored in detail in Section\,\ref{sec:psr_discuss}.

\subsubsection{Ultracool dwarf catalogue}

\label{sec:ucd}

We crossmatched our V-LoTSS sample to the UltraCoolSheet \citep{best_william_m_j_2020_4169085}, a compilation of around 3000 ultracool (spectral types $\sim$M6 and later) dwarf detections. Of these 3000 ultracool dwarfs, approximately 850 are in our survey region. We applied proper motion correction based the information available in the `formula' columns, which is a curated list of the best proper motions from the literature. Only one match was found within 10$''$, and that was to the LOFAR discovered T6.5 brown dwarf Elegast \citep{vedanthambrowndwarf}. It is not surprising that we do not detect other known ultracool dwarfs since the radio emission from most of these objects is expected to be below our sensitivity limit \citep{2012ApJ...746...23M,2016ApJ...818...24K}, except for unusual cases like Elegast \citep{vedanthambrowndwarf}.

\subsubsection{Sloan Digital Sky Survey Data Release 17 catalogue}

Since leakage into Stokes\,V from AGN is the largest potential contaminant of false sources in V-LoTSS, we only searched for extragalactic counterparts to V-LoTSS sources with flux densities $\geqslant 5\sigma_{V}$. We considered a V-LoTSS source to be extragalactic if it had a measured spectroscopic redshift in NASA/IPAC Extragalactic Database (NED) within 1$''$ of the Stokes\,I source position. That condition was satisfied for only for one V-LoTSS source, which is associated with a quasar in the Sloan Digital Sky Survey Data Release 17 \citep[SDSS DR17;][]{2022ApJS..259...35A}. The false-association rate between SDSS DR17 sources with spectroscopic redshifts and $\geqslant 5\sigma_{V}$ V-LoTSS sources within a crossmatching radius of $\leqslant$1$''$ is $<0.003$.  

\subsection{Population characteristics of V-LoTSS and literature comparison}

In total, we have identified 56 circularly-polarised sources that have flux densities $\geqslant5 \sigma_{V}$. Based on their association with a star or pulsar, we have also isolated a further 12 circularly-polarised sources that have $4 \leqslant S_{V}/\sigma_{V} < 5$. We expect $\lesssim$1 of these sources to be a false-association. All object information for the 68 V-LoTSS sources is provided in an electronic table, with the column headings detailed in Appendix\,\ref{sec:appendix_tab}. The handedness of the circularly-polarised emission from V-LoTSS sample is roughly evenly split~-- 40\% and 60\% of sources were detected emitting left-handed and right-handed circularly-polarised light, respectively.

\begin{figure*}
\begin{center}
\includegraphics[scale=0.35]{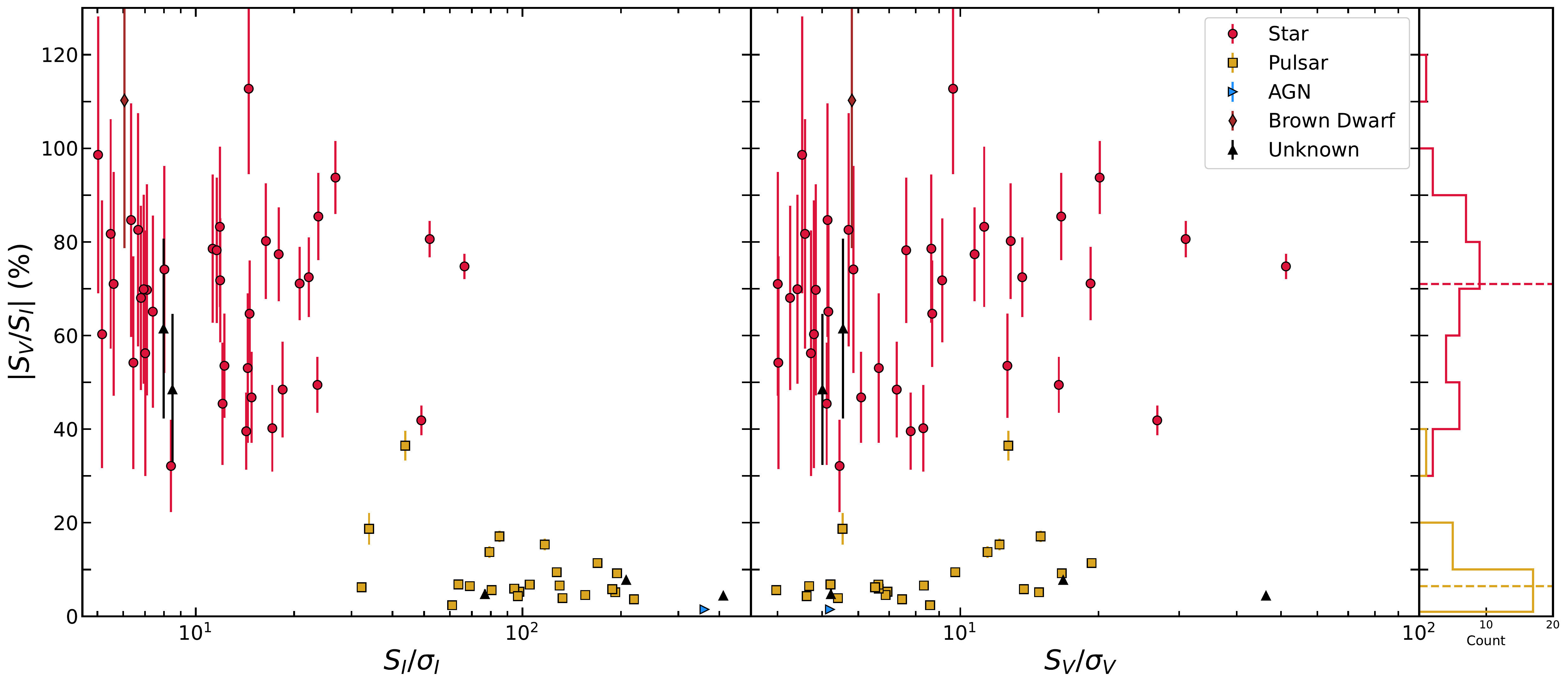}
 \caption{Circularly-polarised fraction of V-LoTSS sources as a function of signal-to-noise ratio in Stokes\,I (left panel) and Stokes\,V (middle panel). We colour each source by its association with a known object class, as outlined in Section\,\ref{sec:targ_assoc}. This includes stars (red circles), pulsars (yellow squares), AGN (blue rightward pointing triangle), or brown dwarf (brown diamond). Sources that have no known association in the literature are represented by black triangles. The histogram (right panel) shows the distribution of the circularly-polarised fraction for sources associated with stars (red line) and pulsars (yellow line), with the median of both of those distributions presented as dashed-lines in their respective colours. The uncertainties on the circularly-polarised fraction are smaller than the size of the symbol unless otherwise plotted.}
\label{fig:circ_pol_frac}
\end{center}
\end{figure*}

The variation of the circularly-polarised fraction with the signal-to-noise ratio in Stokes\,I and V for these 68 sources is shown in Figure\,\ref{fig:circ_pol_frac}. The distribution of the circularly-polarised fraction for V-LoTSS sources associated with known stars or pulsars is presented in the right panel of Figure\,\ref{fig:circ_pol_frac}. V-LoTSS sources are predominately associated with either a pulsar or a star of a various sub-type. These two populations have a significantly different median circularly-polarised fraction. The V-LoTSS sources associated with stars have a median and semi-interquartile (SIQR) circularly-polarised fraction of 71$^{+9}_{-18}$\%. In comparison, the V-LoTSS sources associated with pulsars have a median and SIQR circularly-polarised fraction of 6.5$^{+3.9}_{-1.3}$\%. No pulsar is detected to have a circularly-polarised fraction $>40\%$. We also do not detect a star with a circularly-polarised fraction $<30\%$, suggesting solely based on the circularly-polarised fraction and whether it has an optical counterpart it is possible to identify the likely nature of the unknown sources. The types of stars, pulsars, and the grouping of the unknown sources is discussed in Section\,\ref{sec:discuss}.

For a comparison to the only other wide-field circularly-polarised survey that has been fully published, we note that none of the sources that were detected by \citet{2018MNRAS.478.2835L} fall within the survey area of our current release of V-LoTSS. However, \citet{2018MNRAS.478.2835L} also provided 4$\sigma$ upper-limits on the circularly-polarised fraction of pulsars in their survey region. Only one pulsar with a reported upper limit is located in our survey footprint. In this case, \citet{2018MNRAS.478.2835L} expected the circularly-polarised fraction of PSR\,B0823+26 to be $<13.1\%$. We measure the circularly-polarised fraction of PSR\,B0823+26 to be 2.4$\pm$0.3\%. 

Furthermore, \citet{2018MNRAS.478.2835L} detected a number of artificial satellites in their survey. In contrast, we have not detected any. Since our search for circularly-polarised sources was conducted on 8\,h long exposures, we have washed out any short circularly-polarised emission that was either broadcasted or reflected by an artificial satellite. The artificial satellites detected by \citet{2018MNRAS.478.2835L} were in single 2-min exposures. 

We also do not detect a circularly-polarised stellar system source with a Stokes\,V flux density $>9$\,mJy in the 8\,h long LoTSS exposures. Therefore, it is unsurprising that \citet{2018MNRAS.478.2835L} did not detect the general population of circularly-polarised stars since their survey had a typical sensitivity of $\approx$3\,mJy\,beam$^{-1}$, making V-LoTSS over 20 times more sensitive. However, several of our circularly-polarised stars have been shown to reach $>$\,200\,mJy in Stokes\,V at 144\,MHz on short timescales \citep[e.g.][]{2021A&A...648A..13C}. It appears unlucky that \citet{2018MNRAS.478.2835L} did not catch a southern hemisphere analogue of these stars in outburst.

\subsection{Completeness}

To calculate the completeness of V-LoTSS, we injected point sources into the residual maps and determined our recovery rate with \textsc{Aegean}. The simulated sources were delta-functions convolved with the synthesised beam. Unlike LoTSS-DR2, we did not simulate the impact of ionospheric distortions on source recovery since we are working with 20$''$ resolution images in Stokes\,V. We find V-LoTSS is 50\%, 90\%, and 95\%
complete at 0.3\,mJy, 0.7\,mJy, and 0.9\,mJy, respectively. These completeness values are at slightly higher flux densities than the equivalent Stokes\,I completeness values since we are searching for sources further down the primary beam of single LoTSS pointings, where the noise is higher than if we had mosaicked the data.

\section{Discussion}
\label{sec:discuss}
V-LoTSS is composed of four distinct classes of objects: 1) non-degenerate Galactic circularly-polarised sources; 2) degenerate Galactic circularly-polarised sources; 3) extragalactic circularly-polarised sources; and, 4) unidentified circularly-polarised sources. We discuss each of these classes in turn, and the implications of the source counts of V-LoTSS for a wide-field circularly-polarised survey conducted by the upcoming Square Kilometre Array (SKA). 

\subsection{Galactic non-degenerate circularly-polarised sources}
\label{sec:stars}
\begin{figure*}
\begin{center}
\includegraphics[scale=0.6]{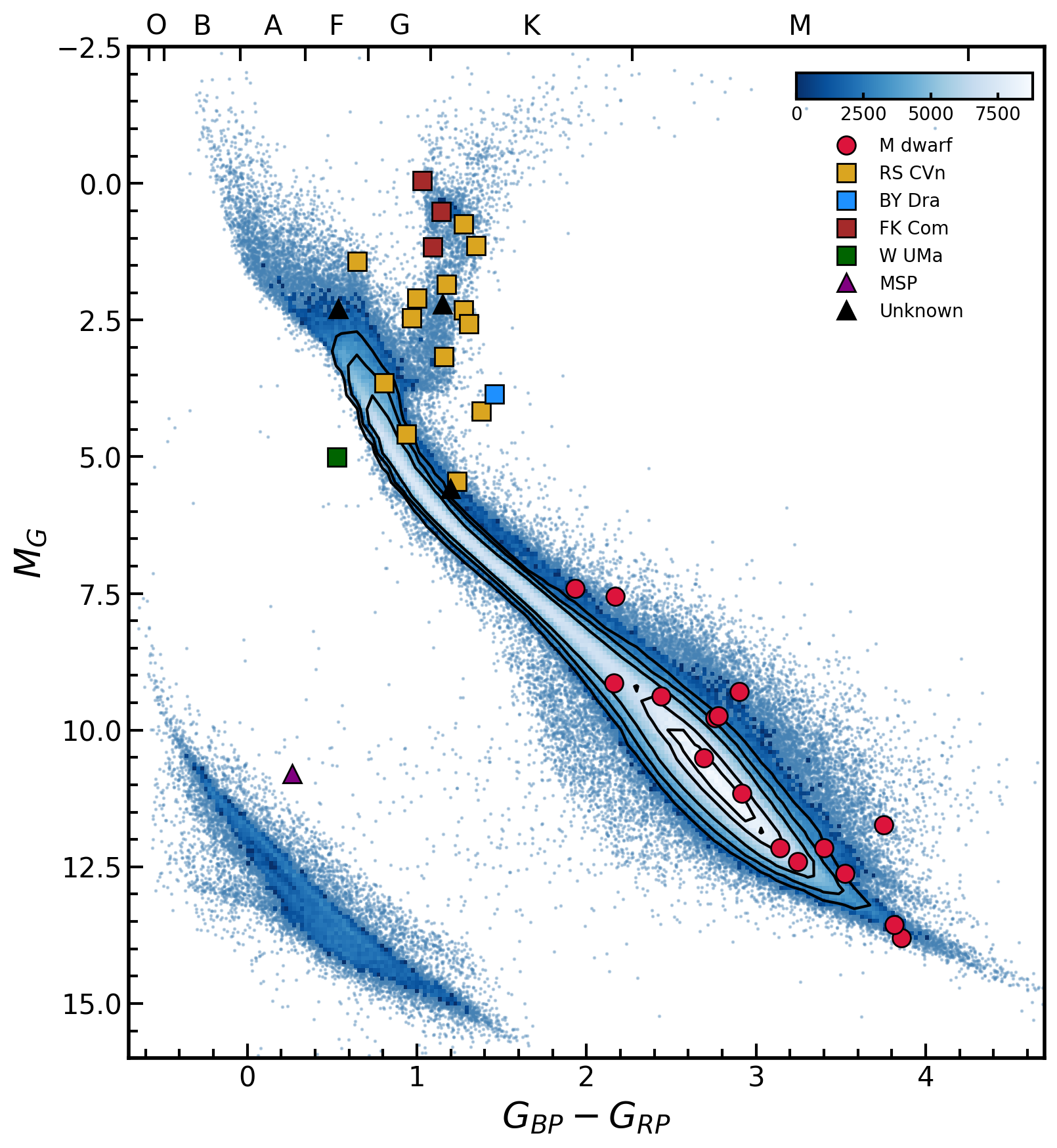}
 \caption{Hertzsprung-Russell diagram of sources in \emph{Gaia} DR3 that are at $\leqslant$150\,pc \citep{2018A&A...616A..10G} and in the V-LoTSS survey footprint, with our Stokes\,V detections with \emph{Gaia} DR3 counterparts overplotted. $M_{G}$ and $G_{BP} - G_{RP}$ represent the absolute magnitude in the \emph{Gaia} $G$-band and colour derived from the blue (BP) and red (RP) parts of the \emph{Gaia} band, respectively. The colour bar in the upper-right corner represents the number of sources in the density plot. The top axis communicates the approximate ranges of different stellar spectral types. The colour of the over-plotted points maps to the source class, as communicated in the legend. The source classes are M dwarfs (red circles), RS Canum Venaticorum (RS\,CVn; yellow squares), BY Draconis (BY\,Dra; blue squares), FK Comae Berenices (FK\,Com; brown squares), and W Ursae Majoris (W\,UMa; green squares) variable stars, and millisecond pulsars (MSP; purple triangle). We plot sources without a known class in the literature as black triangles.} 
\label{fig:hr}
\end{center}
\end{figure*}

The 37 V-LoTSS detections that are associated with Galactic non-degenerate sources are largely composed of two source classes: low-mass dwarfs and extremely chromospherically-active stars. The differentiation into predominantly two source classes is demonstrated in the Hertzsprung-Russell diagram of Figure\,\ref{fig:hr}, where most V-LoTSS sources either follow the M dwarf main-sequence or the giant branch. 

The chromospherically-active stars can be further subdivided into various variable star sub-types, such as RS Canum Venaticorum (RS\,CVn; yellow squares), BY Draconis (BY\,Dra), FK Comae Berenices (FK\,Com), and W Ursae Majoris (W\,UMa) variable stars. In all cases, these stars are either close ($\lesssim$1\,au) young binaries or giants that are rapidly ($\lesssim$2\,d) rotating. All of these stars have luminous ($\gtrsim$10$^{28}$\,erg\,s$^{-1}$) X-ray counterparts. The high-brightness temperature, circularly-polarised radio emission from these objects could plausibly be generated by ECMI via a sub-Alfv\'{e}nic interaction between the binary members or the breakdown of co-rotation \citep{toet21}. However, the exact engine generating the radio emission from these objects is still debated \citep[see e.g.][]{2022ApJ...926L..30V}.

In contrast, the M dwarf sample in V-LoTSS is not dominated by binaries. The V-LoTSS M dwarf sample is composed of stars at all coronal and chromospheric activity levels, as well as slow and fast rotators \citep{CallinghamPopulation}. In particular, it has been suggested that the highly circularly-polarised radio emission from the inactive, slow-rotating M dwarfs in this sample is evidence of a sub-Alfv\'{e}nic interaction between a star and a putative planet \citep{2020NatAs.tmp...34V,CallinghamPopulation,2021ApJ...919L..10P,2022MNRAS.514..675K}. For the more active M dwarfs, the circularly-polarised radio emission could be generated by plasma processes or ECMI via the breakdown of co-rotation \citep{2021A&A...648A..13C}. Also, as noted in Section\,\ref{sec:ucd}, we identify only one known ultracool dwarf in V-LoTSS, which was previously discovered using preliminary V-LoTSS data \citep{vedanthambrowndwarf}.

For the remainder of this section we will focus on the three stars that do not have a clear source class identified in the literature. We aim to determine if these three stars are consistent with the chromospherically-active, M dwarf, or a new stellar sub-group in V-LoTSS.

HD\,220242 is the only isolated, main-sequence star other than an M dwarf that we have detected as highly circularly polarised. It has a spectral class of F5\,V, and does not appear to have a close stellar companion \citep{1997A&AS..126...21N} or an anomalously high X-ray luminosity for its spectral type \citep{2003ApJ...595.1206S}. Furthermore, it does not have strong coronal or chromospheric activity indicators. Based on our sensitivity horizon to M\,dwarfs of $\approx$50\,pc \citep{CallinghamPopulation}, and assuming the mechanism driving the circularly-polarised emission from the M\,dwarfs is universal across the different spectral types, we expect approximately one detection of a F, G, or K dwarf given that we have detected 16 M\,dwarfs \citep{Henry2006}. However, there is a long-period astrometric signal that suggests HD\,220242 could have a distant ($>10$ au) companion that is consistent with the mass of a M or brown dwarf \citep{2019A&A...623A..72K,2022A&A...657A...7K}. Therefore, it is possible that the circularly-polarised emission we are detecting is from this low-mass companion. Further radial-velocity monitoring is required to determine the significance of the astrometric signal. Very long baseline interferometry could also be used to resolve the location of the radio emission to the primary or putative companion if the radio emission extends to gigahertz frequencies. 

TYC\,2834-1385-1 may also be an isolated main-sequence dwarf but limited information is available about the source in the literature to be conclusive about its nature. As shown in Figure\,\ref{fig:hr}, TYC\,2834-1385-1 is consistent with an early K dwarf, and \emph{Gaia} DR3 does not report a significant amount of excess astrometric noise for the source. It is not detected in the ROSAT all-sky survey \citep{2016A&A...588A.103B}, with a three-sigma upper-limit on its soft (0.1-2.4\,keV) X-ray luminosity of $\lesssim$1.4\,$\times$\,10$^{31}$\,erg\,s$^{-1}$. While all other chromospherically-active stars identified in V-LoTSS are detected in the ROSAT all-sky survey, the upper-limit of $\lesssim$10$^{31}$\,erg\,s$^{-1}$ is consistent with the average soft X-ray luminosity of the chromospherically active star sample \citep{2022ApJ...926L..30V}. TYC\,2834-1385-1 does not appear to be optically variable \citep{2004AJ....127.2436W,2018AJ....155...39O}. Perhaps most cogent is the fact that TYC\,2834-1385-1 is located at $\approx$140\,pc. Such a distance is more than double the furthest M\,dwarf V-LoTSS detection, but consistent with the average distance of the chromospherically active star population in V-LoTSS. We suggest that this star is likely an unidentified RS\,CVn or FK\,Com variable. To test if this is the case, a high-resolution optical spectrum is required to determine if the star is a close binary and measure the rotation rate from line broadening. Also, a sensitive X-ray observation would establish whether its soft X-ray luminosity is underluminous for a chromospherically active star.

BD\,+42$^{\circ}$\,2437 has a known soft X-ray counterpart with a high luminosity of $\approx$10$^{31}$ erg\,s$^{-1}$ \citep{2016A&A...588A.103B}, and a fast $\approx25.2$\,d rotation period for a giant \citep{2013AcA....63...53K}. The high X-ray luminosity, location on the giant branch, and fast rotation also suggest BD\,+42$^{\circ}$\,2437 is consistent with the chromospherically-active stars sub-group in V-LoTSS.

\subsection{Galactic degenerate circularly-polarised sources}
\label{sec:psr_discuss}

\begin{figure*}
\begin{center}
\includegraphics[scale=0.4]{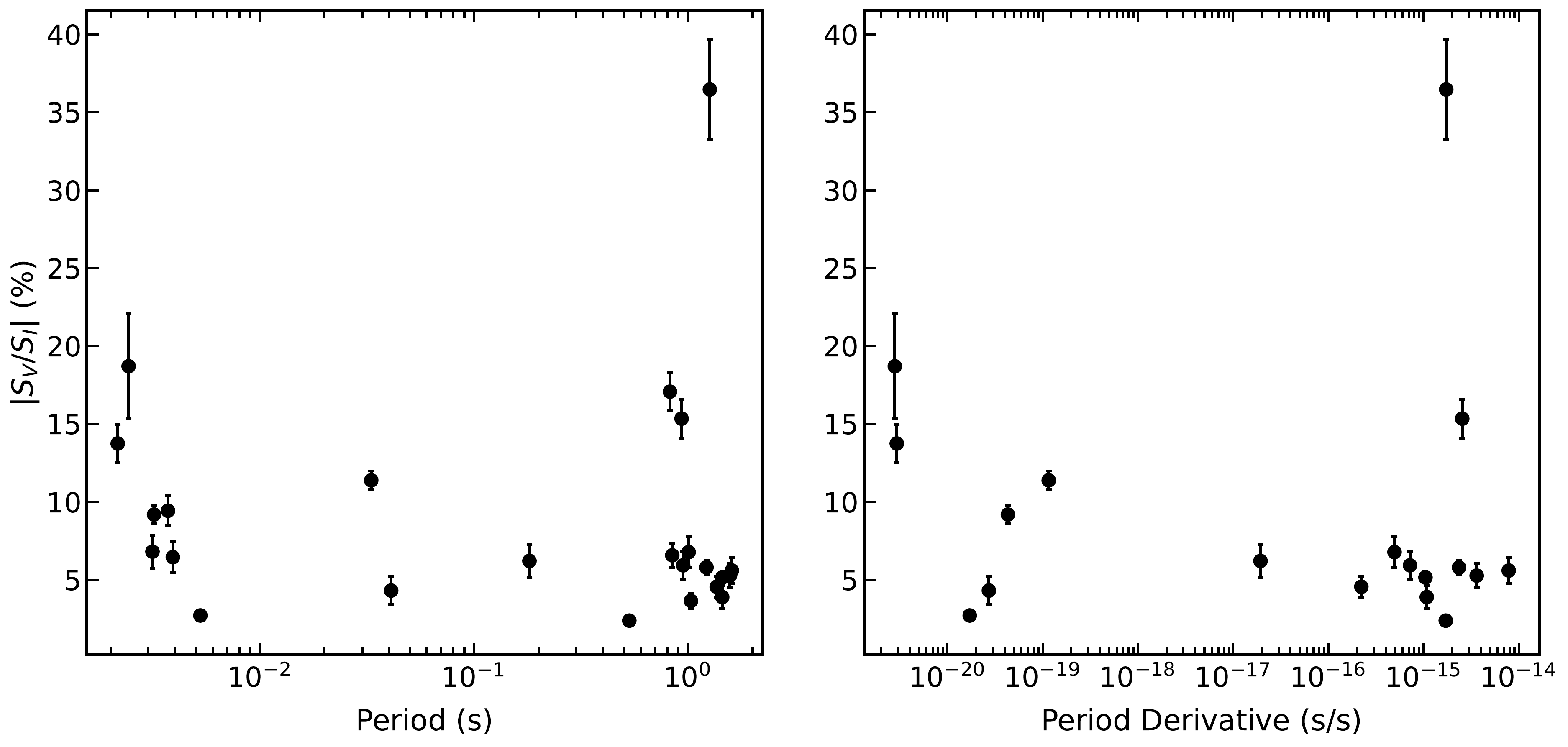}
 \caption{Circularly-polarised fraction of known pulsars as a function of pulse period (left panel) and period derivative (right panel). The uncertainties in period and period derivative are smaller than the symbol sizes. We do not have a measurement of period derivative for five of the detected pulsars. We observe no trend of circularly-polarised fraction with either period or period derivative.}
\label{fig:circ_pol_frac_psr}
\end{center}
\end{figure*}

The pulsars we detect as circularly polarised have periods that span $\approx$2\,ms to 1.6\,s. One-quarter of the circularly-polarised pulsars we have detected are MSPs with period $<$6\,ms, consistent with the underlying high ($>$20$^{\circ}$) Galactic latitude population of these old MSPs \citep{2005AJ....129.1993M}. As shown in Figure\,\ref{fig:circ_pol_frac_psr}, we observe no trend in circularly-polarised fraction with period or period derivative -- MSPs are as circularly polarised as their slower rotating counterparts. 

Only two of the detected pulsars in V-LoTSS have measured circularly-polarised fraction of their average pulse profile around 150\,MHz. We measure a circularly-polarised fraction in our images of $2.4\pm0.3$\,\% and $2.7\pm0.3$\,\% for PSR\,B0823+26 and PSR\,J1012+5307, respectively. In comparison, the circularly-polarised fraction of the average pulse profile of PSR\,B0823+26 and PSR\,J1012+5307 are $6.21\pm0.07$\,\% and $9\pm1$\,\%, respectively \citep{2015A&A...576A..62N}. The relative de-polarisation between the beam-formed and interferometric observations is likely because the interferometric measurements average over all pulse phases. In comparison, the beam-formed observations of \citet{2015A&A...576A..62N} averages only over the on-pulse region of the pulse profile.

\subsection{Extragalactic circularly-polarised sources}

We only detect one circularly-polarised source that is associated with an AGN: GB6\,B1639+5425 (WN\,B1639.4+5425; SDSS\,J164033.57+541943.3). Based on its SDSS properties, GB6\,B1639+5425 is a broad-line quasar at redshift 2.87 \citep{2020ApJS..250....8L}. The source exhibits a peak in its radio spectrum around 1\,GHz, making it a candidate peaked-spectrum source \citep{2017ApJ...836..174C}. However, contemporaneous observations across a wide bandwidth are required to ensure that such a peak is not the product of variability. We note that the source does not display measurable variability between two different LoTSS fields that were observed separated by a day, or between NVSS \citep{Condon1998} and FIRST \citep{1997ApJ...475..479W}.

The circularly-polarised fraction of GB6\,B1639+5425 is somewhat high for an AGN at $1.4\pm0.3\%$, which is consistent with the radio emission from GB6\,B1639+5425 being dominated by a compact pc-scale jet -- as revealed by very long-baseline interferometric (VLBI) observations of the source \citep{2021AJ....161...14P}. 

Our detection of circularly-polarised emission from GB6\,B1639+5425 represents the lowest frequency detection of circularly-polarised emission from an AGN. We also measured the Stokes Q and U properties of GB6\,B1639+5425 to ensure the circularly-polarised signal is not produced by leakage from Stokes Q/U into V. We do not detect linearly polarised emission, meaning that GB6\,B1639+5425 is $<$1\% linearly polarized for rotation measures within the range of $\pm$450 rad\,m$^{-2}$ for channel widths of 97.6\,kHz.

To extract physical insights about the radio jets present in GB6\,B1639+5425 requires us to determine the dominant mechanism that is producing the circularly-polarised radiation \citep{osullivan2013}. With just one measurement at one frequency it is difficult to isolate what is producing the circularly-polarised radiation. For example, it is possible that the circular polarisation is produced by Faraday conversion of linear polarisation \citep{osullivan2013}, synchrotron emission itself \citep{1968ApJ...154..499L}, or even scintillation \citep{2000ApJ...545..798M}. 
A program to contemporaneously observe GB6\,B1639+5425 from $\sim$100 MHz to $\sim$10\,GHz, measuring how the fraction of polarisation changes with frequency and any change in handedness, is required to determine the most likely production mechanism. 

\subsection{Unidentified circularly-polarised sources}

We have five sources that do not have a counterpart in \emph{Gaia} DR3, pulsar, or brown dwarf catalogues. However, based on the circularly-polarised fraction of these sources, we can propose a likely source class association. The unidentified sources with circularly-polarised fractions $<10\%$ are highly likely to be pulsars because, as shown in Figure\,\ref{fig:circ_pol_frac}, almost all detected pulsars have a circularly-polarised fraction $<10\%$. A pulsar classification is especially likely if the source also has a Stokes\,I emission $\gtrsim20$\,mJy in the 8\,h exposures since we do not detect a star with a flux density in Stokes\,I exceeding $\approx$9\,mJy. However, caution is needed when applying this signal-to-noise selection, especially if the exposure time in searching for Stokes\,V emission is small, as stars can burst to $>200$\,mJy on short time scales \citep[e.g.][]{2021A&A...648A..13C}. Pulsar searches are reasonably complete at our survey sensitivity \citep{2018ApJ...866...54T,2020MNRAS.492.5878T,2020MNRAS.491..725M}. Therefore, it is possible that these unidentified pulsars are preferentially MSPs or slowly ($\gtrsim$1\,s) rotating since traditional pulsar searches are more incomplete at those spin periods.

For the sources with a circularly-polarised fraction $>30\%$ but without an optical counterpart in \emph{Gaia} DR3, we suggest these sources are brown dwarf candidates. Similar to the LOFAR discovered brown dwarf Elegast \citep{vedanthambrowndwarf}, deep infrared imaging and spectroscopy is needed to confirm this class suggestion.

To be explicit, we suggest that ILT\,J123927.33+323923.4, ILT\,J132707.56+342337.9, and ILT\,J152745.28+545324.7 are highly likely to be pulsars. In comparison, ILT\,J100804.68+594732.5 and ILT\,J104408.20+341243.6 are brown dwarf candidates. However, we do issue caution in following-up the brown dwarf candidates. We imaged the data on ILT\,J100804.68+594732.5 and ILT\,J104408.20+341243.6 at $\approx$10\,min cadence and do not detect any clear bursts. The radio emission for both sources is at a low-significance for the entire 8\,h observations, similar to what was observed for Elegast \citep{vedanthambrowndwarf}. 

Finally, we can not rule out the possibility that some of these sources could belong to an unknown class of astrophysical objects. However, such a conclusion can only be reached if timing or deep infrared searches at the location of these unidentified sources produce null results. 

\subsection{Source counts and implications for a SKA-Low wide-field survey}

\begin{figure*}
\begin{center}
\includegraphics[scale=0.4]{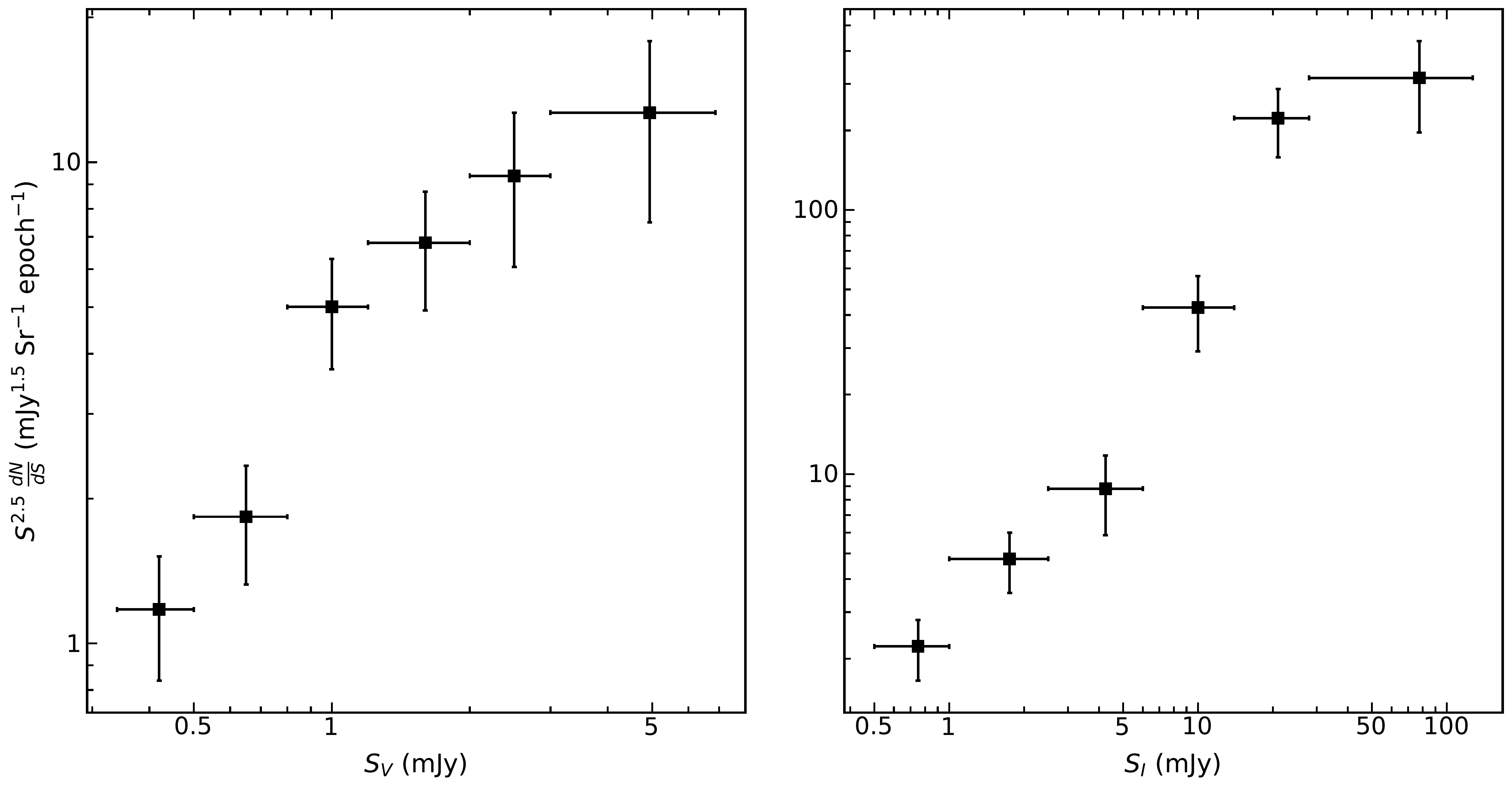}
 \caption{Differential source counts per 8\,h epoch for our Stokes\,V detected sources (left panel) and their Stokes\,I counterparts (right panel). The error bars represent 1-$\sigma$. We have performed a Euclidean Universe normalisation by multiplying the differential counts by flux density $S^{2.5}$. There has been no correction for incompleteness.} 
\label{fig:src_cts}
\end{center}
\end{figure*}

We provide the differential source counts of V-LoTSS in Figure\,\ref{fig:src_cts}. Such information is useful to confirm the flux density at which a survey becomes substantially incomplete and allows us to predict how many circularly-polarised detections we should expect when we have finished V-LoTSS. Figure\,\ref{fig:src_cts} shows that we have not detected a source with a circularly-polarised flux density $>$\,10\,mJy in an 8\,hour synthesised image, and confirms we become
significantly incomplete at circularly-polarised flux densities $<$0.5\,mJy. 

If we assume that circularly-polarised detections are homogeneously distributed across the sky, and that our survey sensitivity will not significantly change with declination, we expect $210\pm40$ detections once V-LoTSS has completed observing the whole northern sky. 

However, this prediction ignores the fact that the detection rate will likely increase towards the Galactic Plane because of the increased density of pulsars in that direction. Since there are $\sim$1,200 known pulsars in the rest of the northern sky yet surveyed \citep{2005AJ....129.1993M}, and we have a detection rate of $\approx$25\% for pulsars in this release of V-LoTSS, we expect $\sim$300 detections of pulsars. However, since the noise in the Galactic Plane will substantially increase relative to observations off the Plane, the true number of pulsars detected at the conclusion of V-LoTSS will likely be closer to $\sim200$ based on an expectation that the noise in the Plane will be on average $\approx$400\,$\mu$Jy. Therefore, we can expect $300 \pm 100$ detections by the completion of V-LoTSS.

Using the trend present in Figure\,\ref{fig:src_cts}, we can also predict the number of detections expected from similar, future low-frequency wide-field surveys. While the current survey strategy of the low-array of the Square Kilometre Array (SKA-Low) is not defined, if SKA-Low conducts a survey with 8\,hour pointings at 148\,MHz with a bandwidth of $\approx$\,50\,MHz, a $5\sigma$ Stokes~V detection will need to be $\approx$50\,$\mu$Jy based on current expected array sensitivity \citep{memo,2019MNRAS.484..648P}. Therefore, we predict that a Southern sky (below declination 0$^{\circ}$) SKA-Low survey could detect 2000$\pm$1000 circularly-polarised sources, of which at least 1000 are expected to be stellar in origin. We have factored the increase in the density of pulsars towards the Galactic Plane into this prediction.

However, since SKA-Low will likely become confusion-limited in total intensity in less than an hour, it appears unlikely that a wide-field survey of 8\,hr pointings will be conducted. Assuming a SKA-Low wide-field survey is conducted with 1\,hr pointings at 148\,MHz with a bandwidth of $\approx$\,50\,MHz, we expect 700$\pm$300 detections provided that a $\geqslant 5\sigma$ Stokes~V detection is $\geqslant 140$\,$\mu$Jy and that the variable circularly-polarised sources identified in 8\,hr epochs are detected at the same rate as those in 1\,hr epochs. 

Finally, note that we performed a Euclidean Universe normalisation for Figure\,\ref{fig:src_cts} out of tradition rather than scientific motivation. Our sensitivity horizon to M\,dwarfs appears to be $\sim$50\,pc, making that population relatively homogeneous in our survey footprint. However, homogeneity and isotropy does not apply to the population of pulsars and chromospherically-active stars. Depending on the source brightness, a spherical or cylindrical geometry should be the correction applied. There is a hint such geometries should be applied as there is more of a knee in the Stokes\,V source counts relative to the Stokes\,I source counts, suggesting there are likely stellar sources in LoTSS-DR2 that do not have Stokes\,V counterparts. Unfortunately, with only 68 sources it is difficult to reliably test which corrections are appropriate. We will conduct such investigation once we produce the final V-LoTSS catalogue of the entire northern sky. 

\section{Conclusions and future outlook}
\label{sec:conc}
We have detailed the production of V-LoTSS -- the most sensitive circularly-polarised radio survey to-date. This data release of V-LoTSS covers $\approx$27\% of the northern sky at a resolution of 20$''$ and with a median noise of 140\,$\mu$Jy\,beam$^{-1}$. We measure the leakage of Stokes\,I into Stokes\,V to be $0.06\pm0.03$\%, and V-LoTSS is 100\% complete at Stokes\,V flux densities $\geq 1$\,mJy.

In total, we identify 68 reliable circularly-polarised low-frequency radio sources. To be deemed a reliable detection, a Stokes\,V source had to be associated with a $\geq$5$\sigma_{I}$ Stokes\,I source and possess a circularly-polarised fraction $\geq1\%$. In general, the Stokes\,V source also had have a flux density $\geq$5$\sigma_{V}$ but we made an exception for 12 Stokes\,V sources with $4 \leqslant S_{V}/\sigma_{V} < 5$ since those sources were also associated with an expected circularly-polarised source class. We expect $\lesssim$1 V-LoTSS source to be a false-association or chance detection. 

We find the sources in V-LoTSS to be composed of four distinct astrophysical classes: stellar systems, pulsars, AGN, or otherwise unidentified in the literature. The stellar systems can be largely separated into chromospherically-active stars, M\,dwarfs, and brown dwarfs. HD\,220242 is the only isolated, quiescent, main-sequence star we have detected that is not an M\,dwarf. However, further observations are needed to confirm that it does not possess a distant low-mass companion and that the radio emission is indeed generated from the F5\,V star. 

We have reliably detected only one AGN in V-LoTSS (GB6\,B1639+5425) -- which has a circularly-polarised fraction of 1.4$\pm$0.3\%. This represents the lowest frequency detection of circular-polarisation from an AGN. The radio emission from this source is dominated by its pc-scale jet. Follow-up observations at higher frequencies is required to differentiate the production mechanism for the circular-polarised radiation.

Based solely on the circularly-polarised fraction, we have also identified pulsar and brown dwarf candidates to be followed-up in radio timing or infrared observations. Finally, once V-LoTSS has covered the entire northern sky, we expect to detect 300$\pm$100 circularly-polarised sources.

We are currently working on improvements for future data releases for V-LoTSS. In particular, we are now producing V-LoTSS images with $\approx$6$''$ resolution and with an improved sensitivity of a factor of $\approx1.2$ over this release. This improvement in sensitivity is due to the downweighting of the difficult-to-calibrate short-baselines of LOFAR. We are also deconvolving the higher-resolution images, facilitating the potential science case of measuring localised circular-polarised emission in resolved sources, such as the hot-spots of radio AGN. The next release of V-LoTSS will likely also involve searching on mosaiced data, potentially recovering constant circularly-polarised, but low-significance, sources. Finally, the analysis of the Stokes Q\,and\,U products of LoTSS (O'Sullivan et al., in prep.), and a search of the Stokes\,V data at shorter $\approx$4\,min cadences (Bloot et al. in prep.), is underway. 

\begin{acknowledgements}
The LOFAR data in this manuscript were processed by the LOFAR Two-Metre Sky Survey (LoTSS) team. This team made use of the LOFAR direction-independent calibration pipeline (\url{https://github.com/lofar-astron/prefactor}), which was deployed by the LOFAR e-infragroup on the Dutch National Grid infrastructure with support of the SURF Co-operative through grants e-infra 160022 e-infra 160152 \citep{2017isgc.confE...2M}. The LoTSS direction dependent calibration and imaging pipeline (\url{http://github.com/mhardcastle/ddf-pipeline/}) was run on compute clusters at Leiden Observatory and the University of Hertfordshire, which are supported by a European Research Council (ERC) Advanced Grant [NEWCLUSTERS-321271] and the UK Science and Technology Funding Council (STFC) [ST/P000096/1]. The Jülich LOFAR Long Term Archive and the German LOFAR network are both coordinated and operated by the J\"{u}lich Supercomputing Centre (JSC), and computing resources on the supercomputer JUWELS at JSC were provided by the Gauss Centre for Supercomputing e.V. (grant CHTB00) through the John von Neumann Institute for Computing (NIC).

JRC thanks the Nederlandse Organisatie voor Wetenschappelijk Onderzoek (NWO) for support via the Talent Programme Veni grant. HK and SB acknowledges funding from the NWO for the project e-MAPS (project number Vi.Vidi.203.093) under the NWO talent scheme VIDI. TWHY acknowledges funding from EOSC Future (Grant Agreement no. 101017536) projects funded by the European Union’s Horizon 2020 research and innovation programme. PNB is grateful for support from the UK STFC via grant ST/V000594/1. MJH acknowledges support from the UK STFC [ST/V000624/1]. MH acknowledges funding from the ERC under the European Union's Horizon 2020 research and innovation programme (grant agreement No 772663). RJvW and RT acknowledges support from the ERC Starting Grant ClusterWeb 804208. GJW gratefully acknowledges the support of an Emeritus Fellowship from The Leverhulme Trust. DJB acknowledges funding from the German Science Foundation DFG, via the Collaborative Research Center SFB1491 "Cosmic Interacting Matters - From Source to Signal". ABonafede, ABotteon, DNH, and CJR acknowledges support from ERC Stg DRANOEL n. 714245 and MIUR FARE grant "SMS". AD acknowledges support by the BMBF Verbundforschung under the grant 05A20STA. KLE is a Jansky Fellow of the National Radio Astronomy Observatory. MHaj acknowledges the MSHE for granting funds for the Polish contribution to the International LOFAR Telescope (MSHE decision no. DIR/WK/2016/2017/05-1) and for maintenance of the LOFAR PL-612 Baldy (MSHE decision no. 59/E-383/SPUB/SP/2019.1). MK acknowledges support from the German Science Foundation DFG, via the Research Unit FOR 5195 "Relativistic Jets in Active Galaxies". MKB acknowledges support from the National Science Centre, Poland under grant no. 2017/26/E/ST9/00216. BM acknowledges support from the UK STFC under grants ST/R00109X/1, ST/R000794/1, and ST/T000295/1. LKM is grateful for support from the UKRI Future Leaders Fellowship (grant MR/T042842/1). DGN acknowledges funding from Conicyt through Fondecyt Postdoctorado (project code 3220195). MPT acknowledges financial support from the Spanish Ministerio de Ciencia e Innovaci\'{o}n (MCIN), the Agencia Estatal de Investigaci\'{o}n (AEI)  through the "Center of Excellence Severo Ochoa" award to the Instituto de Astrofísica de Andalucía (SEV-2017-0709) and through grant PID2020-117404GB-C21 funded by MCIN/AEI/10.13039/501100011033. TPR acknowledges support from the ERC Grant No. 743029 (EASY). AR acknowledges funding from the NWO Aspasia grant (number: 015.016.033). MV acknowledges financial support from the Inter-University Institute for Data Intensive Astronomy (IDIA), a partnership of the University of Cape Town, the University of Pretoria, the University of the Western Cape and the South African Radio Astronomy Observatory, and from the South African Department of Science and Innovation's National Research Foundation under the ISARP RADIOSKY2020 Joint Research Scheme (DSI-NRF Grant Number 113121) and the CSUR HIPPO Project (DSI-NRF Grant Number 121291).

This research has made use of the SIMBAD database, operated at CDS, Strasbourg, France, and NASA's Astrophysics Data System. This work has also made use of TOPCAT \citep{2005ASPC..347...29T}; the \textsc{IPython} package \citep{PER-GRA:2007}; SciPy \citep{scipy}; \textsc{matplotlib}, a \textsc{Python} library for publication quality graphics \citep{Hunter:2007}; \textsc{Astropy}, a community-developed core \textsc{Python} package for astronomy \citep{2013A&A...558A..33A}; and \textsc{NumPy} \citep{van2011numpy}.  
\end{acknowledgements}

%
%

\bibliographystyle{aa.bst}
\bibliography{vlotss.bbl}

\begin{appendix}
\onecolumn

\section{Description of the V-LoTSS catalogue}\label{sec:appendix_tab}

The column numbers, names, and units for the V-LoTSS catalogue are described below in Table\,\ref{tab:append}. The table is available online. All Stokes\,I information is similar to that found in the LoTSS-DR2 catalogue but can differ depending on whether the source is variable \citep{lotss-dr2}. For the few sources that are not present in LoTSS-DR2 because of their variability, we have formed their LoTSS name in line with the convention defined by \citet{lotss-dr2}. We report some \emph{Gaia}\,DR3 measurements \citep{gaiadr3} if the Stokes\,V source has a counterpart in that survey. Similarly, we report some pulsar properties if the source has a counterpart in Australia Telescope National Facility Pulsar Catalogue \citep[PSRCAT; v1.68;][]{2005AJ....129.1993M}.

\begin{longtable}{cccl}
\caption{\label{tab:append} The column headings and descriptions for the V-LoTSS catalogue.}\\
\hline
\hline
Number & Name & Unit & Description \tabularnewline
\hline
1  & VLoTSS\_name  & -- & Name of the source following the naming convention of LoTSS-DR2 \tabularnewline 
2  & Common\_name  & -- & Name of the source in SIMBAD or PSRCAT (if known) \tabularnewline 
3  & RA\_v & degrees & Stokes\,V Right ascension (J2000)  \tabularnewline 
4  & e\_RA\_v & arcsec & rms uncertainty in Stokes\,V RA \tabularnewline 
5  & Dec\_v & degrees & Stokes\,V Declination (J2000) \tabularnewline 
6  & e\_Dec\_v & arcsec & rms uncertainty in Stokes\,V Declination \tabularnewline 
7  & RA\_i & degrees & Stokes\,I Right ascension (J2000)  \tabularnewline 
8  & e\_RA\_i & arcsec & rms uncertainty in Stokes\,I RA \tabularnewline 
9  & Dec\_i & degrees & Stokes\,I Declination (J2000) \tabularnewline 
10  & e\_Dec\_i & arcsec & rms uncertainty in Stokes\,I Declination \tabularnewline 
11 & Total\_flux\_v & mJy & Integrated Stokes\,V flux density of the source \tabularnewline 
12  & e\_Total\_flux\_v & mJy & 1-$\sigma$ uncertainty on the total Stokes\,V flux density \tabularnewline 
13  & Total\_flux\_i & mJy & Integrated Stokes\,I flux density of the source \tabularnewline 
14  & e\_Total\_flux\_i & mJy & 1-$\sigma$ uncertainty on the total Stokes\,I flux density \tabularnewline 
15  & v\_i & \% & Absolute value of the ratio of the Stokes\,V and Stokes\,I total flux density \tabularnewline 
16  & e\_v\_i & \% & 1-$\sigma$ uncertainty in the circularly-polarised fraction \tabularnewline 
17  & local\_rms\_v & mJy\,beam$^{-1}$ & Average background rms value of the source in Stokes\,V \tabularnewline 
18  & local\_rms\_i & mJy\,beam$^{-1}$ & Average background rms value of the source in Stokes\,I \tabularnewline 
19  & field\_name & -- & LoTSS-DR2 field name that contains the brightest detection of the source \tabularnewline 
20  & date\_obs & YYYY-MM-DD & Date of the LOFAR observation for the reported field  \tabularnewline 
21  & dup\_detect & -- & Number of independent detections in different LoTSS fields  \tabularnewline 
22  & designation\_gaia & -- & \emph{Gaia}\,DR3 source designation \tabularnewline 
23  & RA\_gaia & degrees & \emph{Gaia}\,DR3 Right ascension (J2000)  \tabularnewline 
24  & Dec\_gaia & degrees & \emph{Gaia}\,DR3 Declination (J2000)  \tabularnewline 
25  & parallax & milliarcsec & \emph{Gaia}\,DR3 parallax \tabularnewline 
26  & e\_parallax & milliarcsec & Uncertainty in \emph{Gaia}\,DR3 parallax \tabularnewline 
27  & pm & arcsec & \emph{Gaia}\,DR3 proper motion \tabularnewline 
28  & ref\_epoch\_gaia & YYYY.Y & \emph{Gaia}\,DR3 reference epoch \tabularnewline 
29  & phot\_g\_mean\_mag\_gaia & mag & \emph{Gaia}\,DR3 G-band mean magnitude \tabularnewline 
30  & bp\_rp\_gaia & mag & \emph{Gaia}\,DR3 BP-RP colour \tabularnewline 
31  & period\_psr & s & Period of pulsar in PSRCAT \tabularnewline 
32  & e\_period\_psr & s & Uncertainty in the period of pulsar in PSRCAT  \tabularnewline 
33  & period\_deriv\_psr & s\,s$^{-1}$ & Period derivative of pulsar in PSRCAT \tabularnewline 
34  & e\_period\_deriv\_psr & s\,s$^{-1}$ & Uncertainty in the period derivative of pulsar in PSRCAT \tabularnewline 
\hline
\end{longtable}

\end{appendix}
\end{document}